\documentclass[lettersize,journal]{IEEEtran}
\usepackage{cite}
\usepackage{amsmath,amsfonts}
\usepackage{array}
\usepackage{textcomp}
\usepackage{stfloats}
\usepackage{url}
\usepackage{verbatim}
\usepackage{graphicx}
\usepackage{enumerate}
\usepackage{enumitem}
\usepackage{subfigure}
\usepackage{graphicx}
\usepackage{booktabs}
\usepackage[linesnumbered,ruled,vlined,algo2e]{algorithm2e}
\usepackage{todonotes}
\usepackage{amssymb}
\usepackage{xparse}
\newcommand{\ignore}[1]{{}}

\def\BibTeX{{\rm B\kern-.05em{\sc i\kern-.025em b}\kern-.08em
    T\kern-.1667em\lower.7ex\hbox{E}\kern-.125emX}}
\usepackage{balance}

\let\oldnl\nl%
\newcommand{\nonl}{\renewcommand{\nl}{\let\nl\oldnl}}%

\begin{document}
\title{ETCInfer: An Energy-efficient Thermal-aware Cooling-joint Scheduler for LLM Inference in AI Datacenters}

\author{
Rui Lu, 
Rui Ge, 
Huanghuang Liang, 
Xiaobo Zhou,
Dan Wang
\IEEEcompsocitemizethanks{\IEEEcompsocthanksitem 
Rui Lu is with the Department of Computing, The Hong Kong Polytechnic University, Hung Hom 999077, Hong Kong. Emails: (ruilu@polyu.edu.hk)

Dan Wang is with the Division of Environment and Sustainability Academy of Interdisciplinary Studies, Hong Kong University of Science and Technology, Clear Water Bay, Hong Kong. Emails: (wangdan@ust.hk)

Rui Ge, Huanghuang Liang, and Xiaobo Zhou are with the State Key Laboratory of IOTSC, University of Macau, China. Rui Ge and Huanghuang Liang are also affiliated with the School of Computer Science, Wuhan University, China. Emails: (gerui, huanghuangliang, waynexzhou@um.edu.mo)

}
}

\maketitle

\begin{abstract}
{Large language model (LLM) inference in AI datacenters creates a coupled control problem between GPU serving and facility cooling.} Raising ambient temperature setpoints can reduce cooling energy and carbon, but also shrinks thermal headroom, induces GPU throttling, and leads to {Service-Level-Objective (SLO)} violations. 
In this paper, we study {joint cooling--computing control for LLM inference}: {minimizing per-job GPU-plus-cooling energy} while satisfying {thermal safety and latency SLO constraints}. We present ETCInfer, an energy-efficient, thermal-aware scheduler that {selects a pre-job} {Computer Room Air Conditioner (CRAC)} {setpoint and adapts per-GPU frequency and micro-batch size during execution}. 
ETCInfer {builds compact physics-informed control models by calibrating} GPU heat generation, chassis heat dissipation, CRAC power, and {prefill/decode latency relations} from telemetry. These models estimate {hidden thermal states and time-to-throttle}, enabling the scheduler to evaluate energy, temperature, and latency before applying an action.
We formulate {this joint setpoint--frequency--micro-batch control problem} as a partially observable Markov decision process and design ETCAdapter, a learning-based controller that minimizes per-job energy under thermal safety and SLO constraints. 
We implement ETCInfer {as a coordination layer over} typical inference and cluster management stacks. 
Evaluation across real-trace simulation and validation experiments shows that ETCInfer reduces total job energy by up to 33.1\%, thermal throttle exposure by up to 92.9\%, and {keeps SLO violation rates below 0.7\% even at ambient temperatures up to 48$^{\circ}\mathrm{C}$}.
\end{abstract}

\begin{IEEEkeywords}
LLM Inference, Datacenter Scheduling, Energy Efficiency, Thermal Throttle, Parallel Processing, Distributed Inference, GPU Cluster
\end{IEEEkeywords}

\section{Introduction}
Currently, large language models (LLMs) perform remarkably in natural language processing (NLP). Models like GPT from OpenAI and Gemini from Google have demonstrated impressive abilities across various tasks, including question answering, summarization, search, code generation, and induction~\cite{jiang2024megascale, patel2024characterizing}. With rapidly advancing capabilities, LLMs are seeing surging user demand, but this growth also brings substantial computational costs for scaling inference on billion-parameter models. To address these demands, companies are rapidly building new AI datacenters equipped with millions of high-performance GPUs such as Nvidia H100. For example, Microsoft plans to invest \$80 billion in 2025 to expand its AI-enabled datacenters globally~\cite{smith2025golden}. As a result, the energy consumption of AI datacenters has increased significantly, contributing substantially to their carbon footprint. Today, AI facilities are already a visible component of global electricity demand and are projected to reach several percent of worldwide consumption within the next decade.

In AI datacenters, energy consumption is dominated by two closely coupled sources: computational hardware (GPUs) and the cooling systems required to dissipate the resulting heat. High-performance GPUs, such as the NVIDIA H100 SXM, can draw up to 700 W each, amounting to several megawatt-hours annually per GPU~\cite{nvidia_h100_pb11133}. Since most consumed electrical energy ultimately becomes heat in the room, cooling infrastructure is essential for safe operation and hardware reliability, often consuming 30\%--40\% of total datacenter power~\cite{su2019research, wang2024greencooling35}.

\begin{figure*}[t]

    \centering
    \begin{minipage}[t]{0.61\linewidth}
        \centering
        \includegraphics[width=\linewidth]{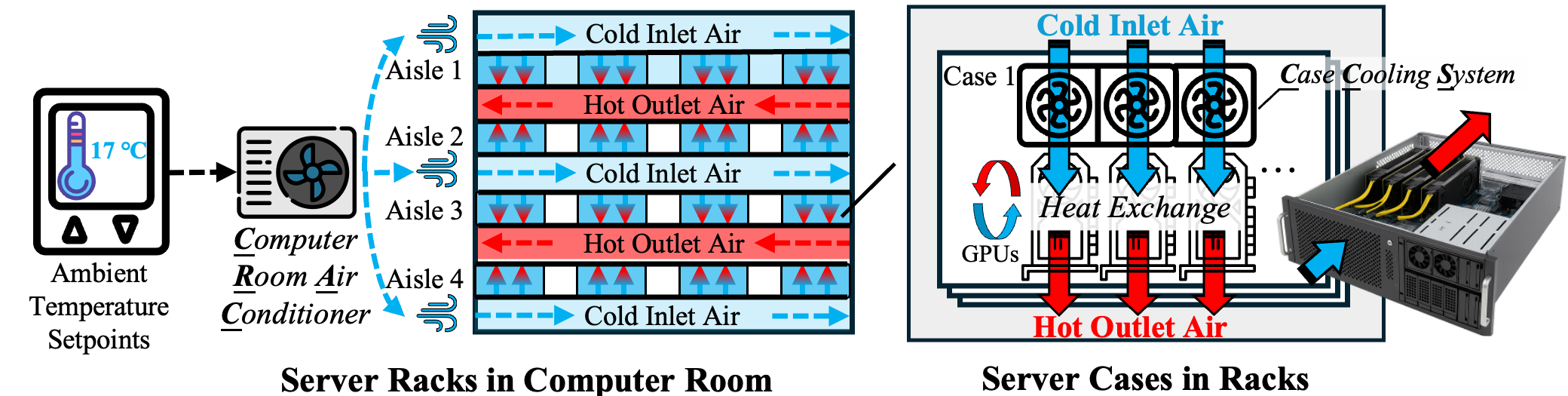}
        \caption{The cooling systems in AI datacenters.}
        \label{fig:server_room}
    \end{minipage}
    \hfill
    \begin{minipage}[t]{0.23\linewidth}
        \centering
        \includegraphics[width=\linewidth]{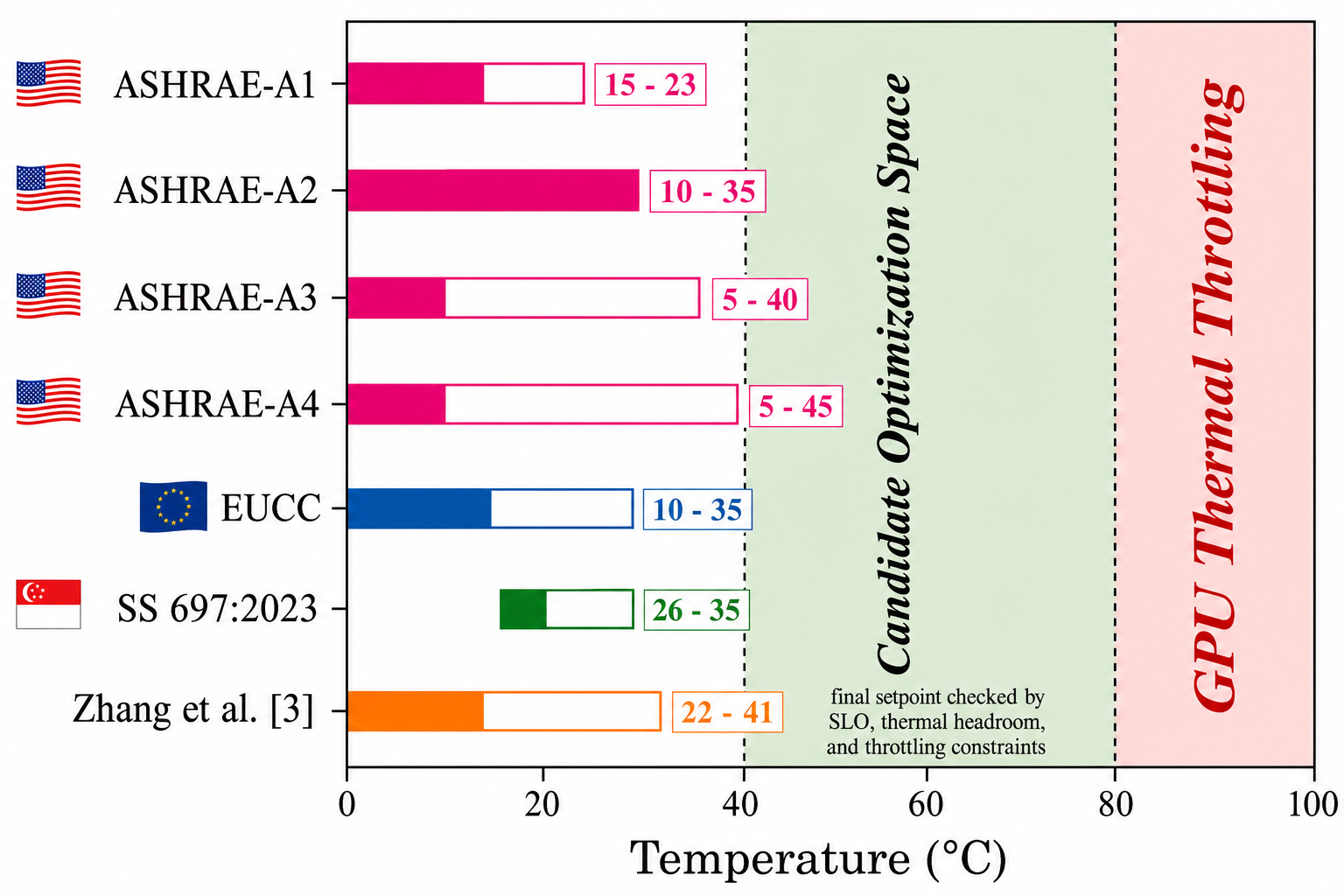}
        \caption{Datacenter guidelines. {Green: candidate setpoints.}}
        \label{fig:guidelines}
    \end{minipage}
    \hfill
    \begin{minipage}[t]{0.145\linewidth}
        \centering
        \includegraphics[width=\linewidth]{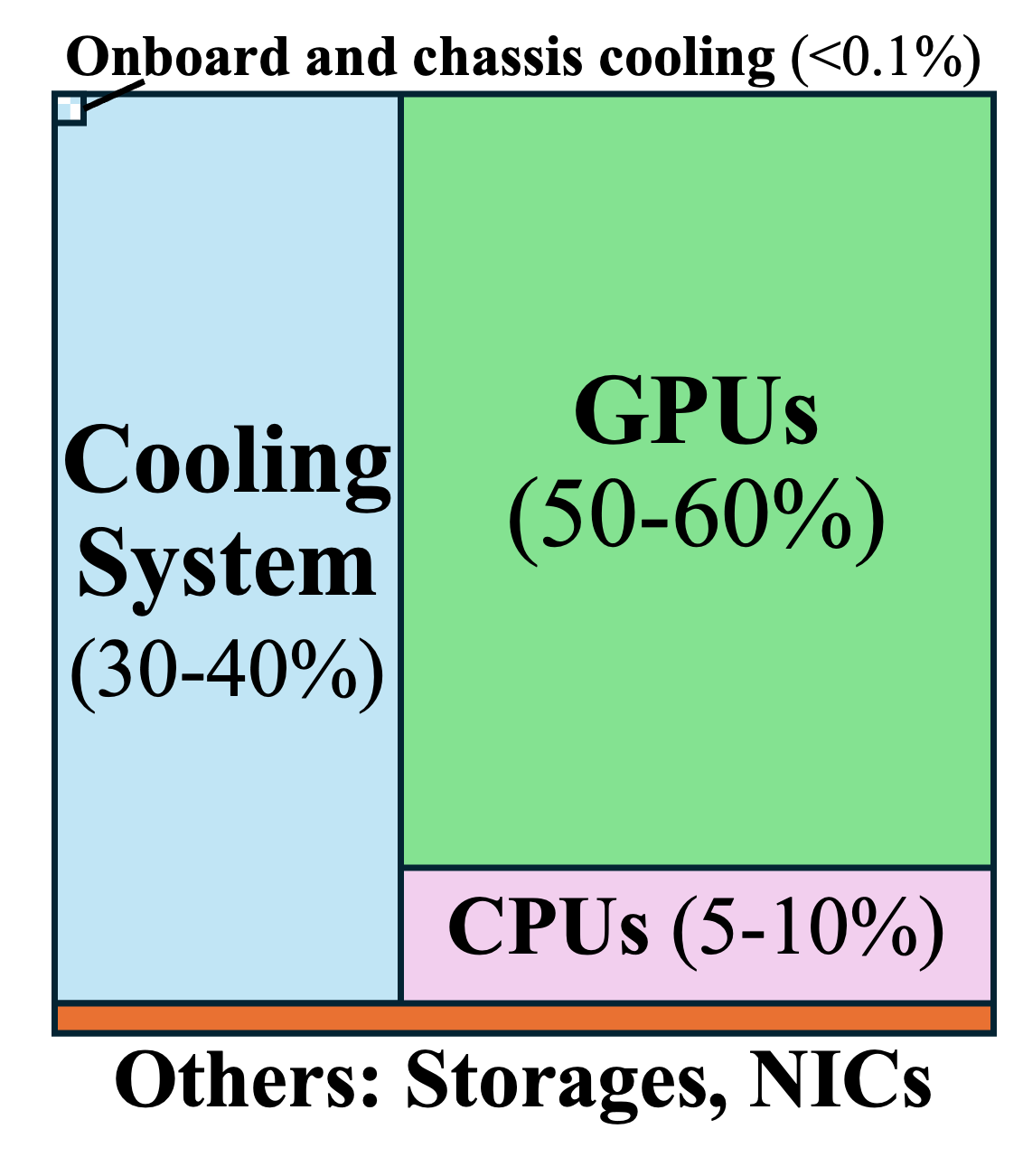}
        \caption{Breakdown of datacenter energy.}
        \label{fig:ratios}
    \end{minipage}
    
\end{figure*}

Raising the ambient temperature in the computer room is therefore an attractive knob to reduce cooling energy and carbon emissions. Recent data-center studies~\cite{zhang2023global41degree} indicate that increasing the supply or ambient temperature to around $41^{\circ}\mathrm{C}$ can cut cooling energy costs by up to $56\%$ compared to conventional settings of $20$--$25^{\circ}\mathrm{C}$. Correspondingly, several standards have been established to guide optimal cooling configurations~\cite{ASHRAE_TC9_9_2021, EUCoC-DC-2024, Acton2025DataCentreEfficiency}, {as summarized in Fig.~\ref{fig:guidelines}}. Singapore recommends $28$--$32^{\circ}\mathrm{C}$ for Level-4 datacenters~\cite{ss697_2023}, while the EU Code of Conduct allows higher operating ranges under appropriate reliability controls~\cite{EUCoC-DC-2024, Acton2025DataCentreEfficiency}. Consequently, operating AI datacenters at $35$--$41^{\circ}\mathrm{C}$ has the potential to yield substantial long-term energy and carbon savings.

However, operating LLM inference at warmer, energy-efficient ambient setpoints introduces coupled thermal and performance challenges. High GPU throughput requires substantial power, generating massive heat. Cooling systems are typically designed for moderate ambient temperatures (around $20$--$30^{\circ}\mathrm{C}$) with adequate inlet airflow. As the ambient temperature rises, the temperature gradient from chip to air narrows, reducing cooling effectiveness and accumulating heat. When junction temperature approaches the limit, the GPU triggers \textit{thermal throttling}~\cite{kim2021ztt, stojkovic2025tapas}, reducing clock frequency and voltage to protect itself, which directly lowers throughput until the temperature returns to a safe range. For heavy LLM inference jobs, repeated or extended throttling can severely impact time-to-first-token (TTFT), time-per-output-token (TPOT), and effective throughput, and accelerate device aging.

Existing work on AI datacenter energy mainly optimizes compute or cooling separately. On the compute side, DVFS and model compression lower GPU energy for LLM workloads~\cite{lai2013latency, xiao2023smoothquant, jin2024comprehensive, tian2024greenllm}. On the cooling side, CRAC-oriented studies raise setpoints and refine cooling design to reduce cooling energy and emissions~\cite{zhang2023global41degree, ASHRAE_TC9_9_2021, EUCoC-DC-2024, Acton2025DataCentreEfficiency, ss697_2023}. At the serving layer, thermal-aware resource provisioning and schedulers such as TEAP and TAPAS~\cite{zhang2025thermal, stojkovic2025tapas} emphasize host allocation, batching, and placement but do not jointly control facility-side setpoints with per-GPU thermal states. {In ETCInfer, rack placement and airflow direction are treated as sources of heterogeneous inlet temperature and cooling delay, not as runtime control actions.} {These techniques provide useful building blocks, but their objectives remain separated across serving, device power, and facility cooling.} As a result, current systems do not jointly optimize ambient setpoint, per-GPU heat dynamics, and job SLOs, leaving the design space where GPU frequency, micro-batch size, and room setpoint interact, leading to three coupled challenges:

\textbf{Challenge 1 (Thermal–performance modeling).}
A thermal prediction model is necessary, which tightly links LLM latency to GPU temperature, especially near the throttle point. As the frequency drops to protect the device, job latency would extend. The key challenge is to predict time-to-throttle and performance from real-time sensor data.

\textbf{Challenge 2 (Joint computing–cooling energy modeling).}
A joint energy model on both GPUs and CRACs is necessary. Raising ambient setpoints reduces cooling power but increases execution time under throttling, canceling savings. The challenge is to couple dynamic GPU heat generation and dissipation so we can quantify {GPU-plus-CRAC per-job energy} under different control actions.

\textbf{Challenge 3 (SLO-aware thermal scheduling).}
An SLO-aware scheduler is necessary that selects and operates at warm ambient setpoints. It also allocates jobs, chooses micro-batch sizes and GPU frequency, and avoids throttling using incomplete thermal information, maintaining SLO guarantees. {The objective is not to maximize the setpoint, but to find a safe operating point where cooling savings do not cause throttling-induced latency violations.}

In this paper, we propose \textit{ETCInfer}, an energy-efficient, thermal-aware cooling-joint scheduler for LLM inference in AI datacenters. {ETCInfer serves as a coordination layer across LLM serving, GPU control, and room cooling, rather than replacing the serving engine, cluster manager, GPU driver, or CRAC controller.}
First, we develop a physics-informed modeling framework that couples GPU heat generation, {air-side heat dissipation}, {room-level cooling}, and {LLM workload latency} to expose per-GPU thermal states, {time-to-throttle}, and SLO forecasts.
Second, we formulate the {joint control of ambient setpoint, GPU frequency, and micro-batch size} as a partially observable decision problem and design ETCAdapter to optimize per-job energy under SLO and thermal-safety constraints.
{ETCInfer combines a joint cooling--computing formulation, physics-informed hidden-state estimation, and online integration with LLM serving controls.}
Third, we implement and evaluate it using trace-driven experiments in real-trace CFD simulation and validation experiments. 
Results show that ETCInfer reduces total job energy by up to 33.1\%, shrinks thermal throttle exposure by up to 92.9\%, and keeps SLO violation rates below 0.7\% even at ambient temperatures up to 48$^{\circ}\mathrm{C}$. 
{We also release the ETCInfer source code to support reproducibility.}
The contributions of this paper can be summarized as follows:

\noindent $~\bullet$ We formulate {joint LLM scheduling and ambient temperature control} as a partially observable decision problem whose actions include {CRAC setpoint}, {GPU frequency}, and {micro-batch size}.

\noindent $~\bullet$ We develop physics-informed control models that couple heat generation, {air-side heat dissipation}, {CRAC power}, and {workload latency} to estimate hidden thermal states, {time-to-throttle}, and SLO risk.

\noindent $~\bullet$ We design ETCAdapter, an online controller that adapts {frequency and micro-batch size} from telemetry and learned latent dynamics while enforcing SLO and thermal-safety constraints.

\noindent $~\bullet$ We implement ETCInfer on {typical LLM inference frameworks} and integrate it into AI datacenter stacks.

\noindent $~\bullet$ Our experiments show ETCInfer {significantly reduces energy and throttling while preserving strict SLOs}.

\section{Background}
\subsection{Thermal Challenges in Modern GPUs.}
Heat generation in scaled semiconductors inside the GPU brings challenges. GPUs convert nearly all electrical power into heat as billions of switching elements toggle at high frequency. Heat flux rises with core frequency and voltage, both kept high to sustain throughput for compute-dense LLM workloads. Advanced semiconductor scaling further amplifies heat density, which increases thermal stress in future GPUs. 
Recent studies show that smaller transistors possess lower thermal mass and fewer heat-dissipation pathways, causing faster temperature accumulation and larger performance loss. 
For example, 5nm GAAFETs trap more heat than 7nm FinFETs~\cite{chhabria2019impact}, while 7nm exhibit ~12K self-heating, equivalent to 5nm structures reaching ~17K. This 5K rise significantly affects performance, increasing gate delay by up to 39\% at 5nm versus 25\% at 7nm, intensifying leakage currents. 

\textbf{Thermal Throttle of GPUs.}
When heat generation exceeds cooling capacity, junction temperature rises, eventually triggering thermal throttling. This reduces voltage and frequency, limiting GPU performance and throughput, especially for compute-intensive LLM inference. Prolonged throttling accelerates GPU aging, increasing risks like electromigration and joint fatigue. Studies show higher temperatures correlate with higher error rates and earlier failures~\cite{ostrouchov2020gpu}. Avoiding throttling improves throughput, energy efficiency, and hardware lifespan.

\textbf{Dynamic Voltage and Frequency Scaling (DVFS) on GPUs.}
DVFS is a common technique for managing GPU power and thermal behavior~\cite{Guerreiro2019Modeling, Wang2022Energy, Nabavinejad2022Coordinated}{, following earlier GPU power/performance and GPGPU power-modeling work~\cite{hong2010integrated,leng2013gpuwattch}}. It adjusts supply voltage and operating frequency to reduce heat generation at the source, lowering the likelihood of thermal throttling. On NVIDIA GPUs, operators can adjust core and memory clocks, voltages, and board power caps through NVML. By balancing power and performance, effective DVFS policies reduce unnecessary energy use and mitigate throttling, improving resource efficiency in AI datacenters.

\subsection{Heat Dissipation in Datacenters} 
The heat dissipation for GPUs of a datacenter is shown in Fig.~\ref{fig:server_room}. 
Typically, servers host multiple identical GPUs (e.g., 6–8 cards) within a single server chassis. Several chassis are mounted in a rack, and racks are arranged in rows. Each chassis and GPU uses front-to-back fans that draw cold air from the cold aisle and exhaust hot air into the hot aisle.
Most facilities implement hot-aisle/cold-aisle containment. {Computer Room Air Conditioner (CRAC)} units ingest hot-aisle return air, remove heat, and deliver conditioned supply air to the cold aisles, closing the thermal loop in the datacenter. 

The fans on chassis and GPUs, typically are supplied by chassis vendors (e.g., Supermicro) and GPU vendors (e.g., NVIDIA) and are designed for ambient temperatures of 20–30$~^\circ$C. Each fan consumes only a few watts, which is negligible compared to GPU power, so their energy cost is often ignored, and they are assumed to operate at maximum RPM. However, their effectiveness decreases as ambient temperature rises~\cite{Zapater2015Leakage, Yao2015Adaptive}.

\textbf{Cooling Guidelines for Datacenters.} In datacenters, CRAC systems (approximately 35\%) and servers/GPUs (approximately 55\%) account for most energy usage~\cite{riu2024loadgrowthE3}, making ambient temperature a key control parameter. Lower setpoints increase onboard and chassis cooling capacity and reduce GPU thermal throttling, but excessive cooling wastes energy. Industry guidelines in Fig.~\ref{fig:ratios} recommend fixed ambient ranges to balance reliability and efficiency: ASHRAE suggests 27$~^\circ$C, while the EU allows up to 35$~^\circ$C~\cite{ASHRAE_TC9_9_2021, EUCoC-DC-2024}. Recent studies indicate that raising ambient temperatures toward 41$~^\circ$C can further save energy with minimal impact on hardware lifespan~\cite{zhang2023global41degree}.
Efficient cooling reduces energy use and also carbon emissions, enabling AI datacenters to support large-scale LLM workloads with lower environmental impact.

\begin{figure*}[t]
    \centering
     \includegraphics[width=0.775\linewidth]{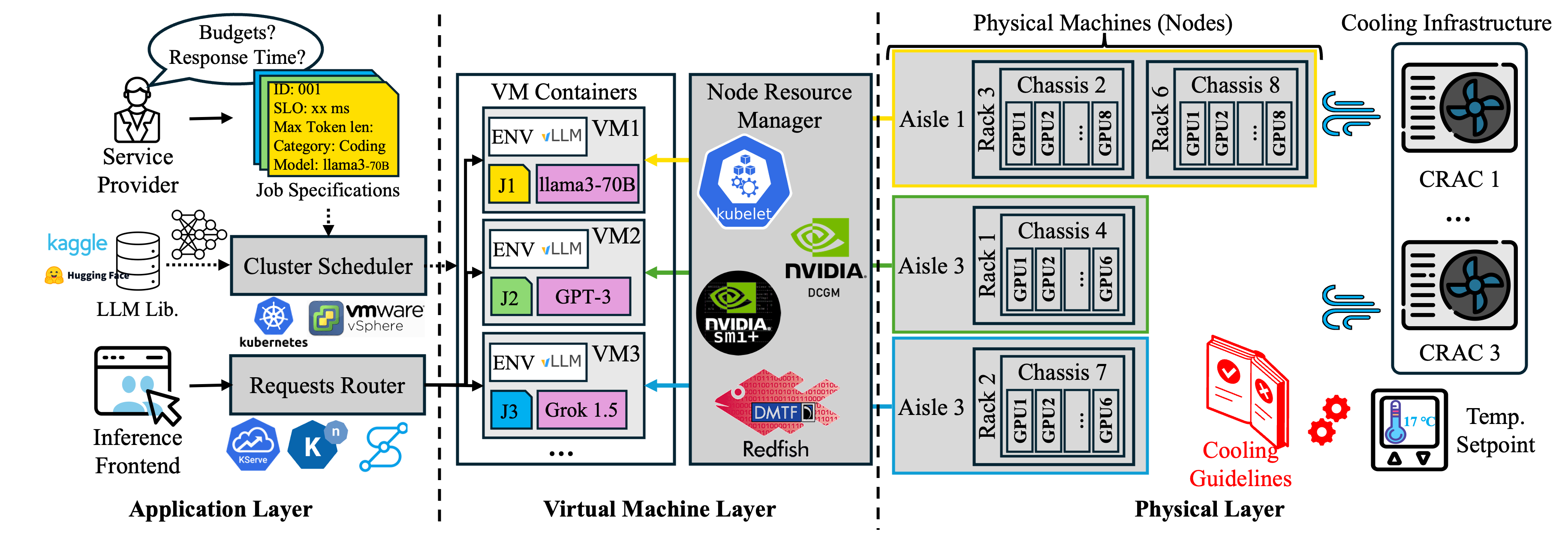}
     
    \caption{{General multi-service LLM inference workflow in an AI datacenter. The figure illustrates the deployment architecture rather than the exact model set used in evaluation.}}
    \label{fig:ex_inf_workflow}
    
\end{figure*}

\subsection{Scheduling of LLM Inference}
LLM inference at datacenter scale requires a scheduler that places requests on heterogeneous GPUs, coordinates batching and model parallelism, and manages memory-intensive state such as KV caches under time-varying load{~\cite{li2023vllm,nvidia_tensorrtllm_repo,Jiang2025Efficient}}{, including prefill/decode and chunked-prefill scheduling systems~\cite{yu2022orca,zhong2024distserve,agrawal2024sarathi}}. The scheduler must balance user experience, cost, and infrastructure limits such as memory and bandwidth. 
However, most production schedulers are \emph{thermal-agnostic}: they abstract GPUs as virtual compute units and ignore per-device physical context such as inlet temperature, fan speed, and thermal headroom. 
At the application layer, platforms schedule \emph{virtual} resources, {such as containers and Pods in Kubernetes or Ray}, while cluster managers that place \emph{physical} GPUs, {such as Kubernetes with the NVIDIA device plugin, SLURM, or LSF}, usually lack feedback loops from {CRACs and device telemetry for per-GPU thermal awareness}. As a result, placement and scaling decisions are driven mainly by software metrics such as queue length, latency, and utilization rather than actual thermal conditions.

\textbf{Thermal Prediction–Driven Thermal-Aware Scheduling.} 
While the energy/carbon-focused scheduler reduces power and emissions, it cannot prevent performance drops when GPUs approach thermal limits. If cooling is slow or inlet temperatures are high, die temperature rises, triggering thermal throttling, which reduces throughput and risks SLO violations. LLM inference amplifies this risk due to long, bursty decode phases that keep power high even at moderate utilization. {Heterogeneous rack placement and airflow create uneven thermal conditions across replicas.} Prolonged throttling accelerates wear, increases error rates, and threatens hardware lifespan and service reliability. These challenges motivate a thermal-aware scheduler that treats temperature and time-to-throttle as key signals, guiding placement, batching, and DVFS adjustments, while coordinating with facility controls to maintain throughput and stability under real heat constraints.

\section{Scheduler Overview}
\subsection{Design Overview}
Modern LLM inference in AI datacenters follows a structured pipeline involving multiple schedulers and runtime engines, as shown in Fig.~\ref{fig:ex_inf_workflow}. {This figure is a general multi-service deployment view: different Pods or virtual services may host different LLM services, while the concrete model choices used in evaluation are specified in Section~\ref{subsec:setup}.}
Before execution, the \textit{Service Provider} defines each job class with detailed specifications, including {TTFT and TPOT SLOs}, input and output token limits, job categories such as code completion or dialog, and {the target model family or service variant, e.g., Llama-3.1-Instruct}. 
The \textit{Cluster Scheduler}, typically implemented on Kubernetes or VMware, interprets these specifications, creates Pods, retrieves models from registries, and prepares the runtime environment, such as vLLM. During inference, it manages auto-scaling, load balancing, and placement. 
The \textit{Node Resource Manager} assigns Pods to physical nodes, exposes hardware states, and enforces resource isolation. Incoming user queries arrive through the Inference Frontend, where the \textit{Requests Router} groups and dispatches them to Pods based on job category, prompt length, and model choice. Cooling infrastructure maintains chassis temperature following site guidelines.

\begin{figure}[t]
    \centering
     \includegraphics[width=0.975\linewidth]{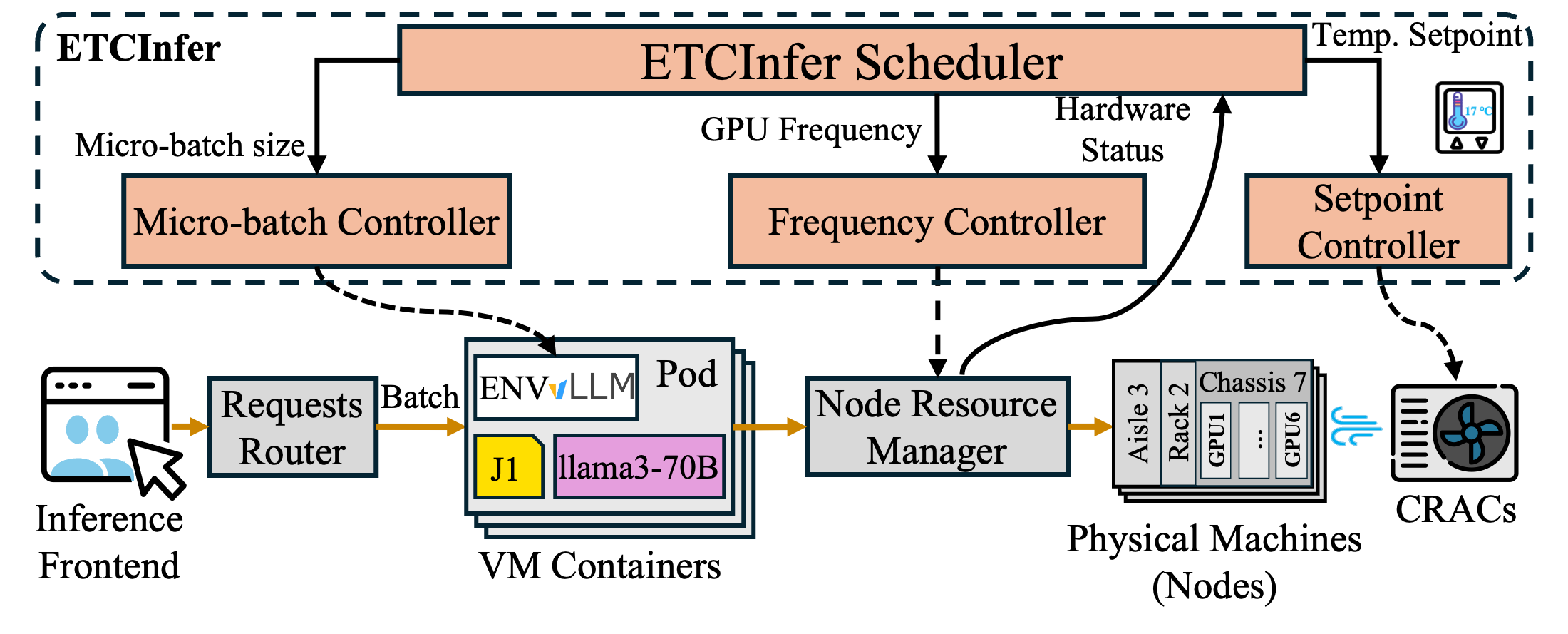}
     
    \caption{Workflow of ETCInfer Scheduler.}
    \label{fig:etcinfer_sch}
    
\end{figure}

In this paper, we design an energy-efficient, thermal-aware cooling-joint Scheduler, \textit{ETCInfer} (Fig.~\ref{fig:etcinfer_sch}), which raises ambient temperature to reduce cooling energy while preserving service quality and hardware lifetime. 
We build a {thermal-state model} linking GPU heat generation and dissipation to latency and energy. ETCInfer coordinates three controls: 
(1) \textit{{Micro-batch control in serving Pods at the virtual machine layer}}, which shapes GPU load, heat, queuing, and SLOs {because a larger micro-batch improves utilization and amortizes overhead but also increases per-step work, power, heat, and latency risk}, 
(2) \textit{{GPU-frequency control on physical machines}} via DVFS, which tunes power and throughput under thermal and SLO limits, 
and (3) \textit{{CRAC setpoint control in the computer room}}, which adjusts ambient temperature within safety margins. 
These jointly predict and schedule thermal dynamics to meet service objectives. {ETCInfer supplies coordinated control decisions to these layers, but it does not replace vLLM batching logic, Kubernetes/KServe placement and routing, GPU driver enforcement, or the facility CRAC controller.}

\noindent\textbf{Assumptions. }
Our scheduling makes several practical assumptions: 
(1) Ambient setpoint changes take effect with delays, so {chassis inlet temperature varies with airflow, rack position, and activity}. 
(2) Chassis fans operate at maximum speed with negligible, constant power, and are excluded from optimization. 
(3) Each GPU follows a driver-level DVFS voltage–frequency curve, and we tune only frequency while voltage adjusts automatically. 
(4) Each inference job is routed to a Pod via the Requests Router and shares stable properties such as SLOs and input length.

\subsection{Scheduling Objectives}
Let $\mathcal{G}_{\mathbf{j}}$ be the set of GPUs of the Pod assigned to job $\mathbf{j}$.
Let $\mathcal{B}$ represent the inference batch with size $N_{batch}$ from the Requests Router and $\mathcal{I}=\{0,1,\dots, I_\mathbf{j}\}$ index control and measurement instants.
For GPU $x\in\mathcal{G}_{\mathbf{j}}$ and time $i\in\mathcal{I}$, sensors provide observation
$\mathbf{\bar o}_i^x=(\bar P_i^x,\bar u_i^x,\bar T_i^x, \bar{T}^x_{in,i})$,  including GPU power $\bar{P}^x_i$, GPU utilization $\bar{u}^x_i$, and GPU core temperature $\bar{T}^x_i$, chassis inlet temperature $\bar{T}^x_{in,i}$.\footnote{In this paper, $\bar{\cdot}$ denotes sensor readings, $\tilde{\cdot}$ denotes physics-based estimates, and $\hat{\cdot}$ denotes learned predictions. For example, $\bar T^x_i$ is sensed temperature, $\tilde T^x_i$ is physics-estimated temperature, and $\hat T^x_i$ is model-predicted temperature.
}

\noindent\textbf{Scheduling Space.}
1) Ambient setpoint ${T_{\mathrm{set}}}\in [T_{\min},T_{\max}] $ is an initialization action before the job begins. It can only be configured, but cannot be adjusted in real-time.
2) Per-GPU frequencies {$f_i^x\in[f_{\min},f_{\max}]$} for all $x\in\mathcal{G}_{\mathbf{j}}$ and $i\in\mathcal{I}$.
3) Each micro-batch allocated in GPU $x$ has size {$N_{mic}^x$} and it must satisfy the SLO (${\mathrm{SLO}_{\mathbf{j}}}$) for the job .
The size of the micro-batch satisfies that $N_{batch} = \sum_{x\in \mathcal{G}_{\mathbf{j}}} N_{mic}^x$. 
{A larger $N_{mic}^x$ assigns more token work to GPU $x$ in one serving step. ETCInfer increases it when SLO slack and thermal headroom are sufficient, and decreases it when predicted temperature, throttling risk, or latency risk rises.}
The three controls operate on different time scales: {setpoint is selected before execution because cooling reacts slowly}, while {frequency and micro-batch size are adapted during execution because they directly change heat generation, throughput, queueing, and SLO risk}.
In summary, the actions at time $i$ could be represented by
\begin{equation}
    \mathbf{a}_0 = T_{\mathrm{set}}, \quad \mathbf{a}_i = \{f_i^x, N_{mic}^x\} \mathrm{\ for\ }i > 0.
\end{equation}

\noindent\textbf{Models and constraints.}
Let $L_{\mathbf{j}}(\cdot)$ denote the latency metrics, e.g., TTFT and TPOT, and let $T_{\mathrm{thr}}^x$ be the throttle threshold. The scheduling problem for minimizing the overall job energy consumption $E(\mathbf{j},T_{\mathrm{set}})$ can be formulated as:
\begin{align}
\min
\quad & {E(\mathbf{j},T_{\mathrm{set}})} \notag\\
\text{s.t.}\quad 
& L_\mathbf{j}\!\left(\{f_i^x\}, N_{mic}^x\right)
   \le \mathrm{SLO}_{\mathbf{j}},  \notag\\
& \bar T_i^x \le T_{\mathrm{thr}}^x, \notag\\
& \forall f_i^x\in[f_{\min},f_{\max}],\ \forall\, x\in\mathcal{G}_{\mathbf{j}},\ \forall\, i\in\mathcal{I}. \notag
\end{align}
{Here, $\bar T_i^x$ is a sensed safety variable at runtime rather than a control variable. ETCInfer acts on $T_{\mathrm{set}}$, $f_i^x$, and $N_{mic}^x$ to keep future sensed temperatures below the throttle threshold.}

\noindent\textbf{Overall Job Energy Consumption.}
To compute the energy of LLM inference job $\mathbf{j}$ on its allocated GPUs and room cooling, let $\Delta t_{i} = t_{i} - t_{i-1}$. The {interval energy over period $i \in \mathcal{I}$} is
\begin{equation}
\Delta E_{i}(\mathbf{j},T_{\mathrm{set}}) = w_{i}\,\tilde P_{\mathrm{CRAC},i-1}(T_{\mathrm{set}})\,\Delta t_{i} + \sum_{x\in\mathcal{G}_{\mathbf{j}}} \bar P^{x}_{i-1}\,\Delta t_{i},
\label{eq:interval_energy}
\end{equation}
where ${\tilde P_{\mathrm{CRAC},i-1}(T_{\mathrm{set}})}$ denotes the estimated electric power consumed by the room cooling system at the selected setpoint and IT heat load, and $w_{i}\in[0,1]$ represents the fraction of shared CRAC power allocated to cooling, computed as $w_{i} = \sum_{x\in\mathcal{G}_{\mathbf{j}}} \bar P^{x}_{i-1} / P_{i-1}^{\mathrm{tot}}$. For a dedicated testbed, set $w_{i}=1$. The total energy to complete job $\mathbf{j}$ at setpoint $T_{\mathrm{set}}$ is
\begin{equation}
E(\mathbf{j},T_{\mathrm{set}}) \;=\; \sum_{i=0}^{I_\mathbf{j}} \Delta E_{i}(\mathbf{j},T_{\mathrm{set}}).
\label{eq:job_energy}
\end{equation}

\subsection{Scheduling Problem as POMDP}
We model the energy-efficient, thermal-aware cooling-joint LLM inference scheduling problem as a \textbf{Partially Observable Markov Decision Process} (POMDP), {which handles control with hidden stochastic states and incomplete observations~\cite{kaelbling1998planning}}. {Scheduling is sequential: each interval-$i$ action changes throughput and power, and affects future die temperature, time-to-throttle, SLO slack, and CRAC energy via thermal inertia.} {Thus, greedy interval control or fixed thresholds may save immediate energy yet trigger later throttling or SLO violations.} {With a fully observed thermal--workload state, the problem becomes an MDP whose transition depends on the current state, action, and stochastic workload.}

ETCInfer uses a POMDP because GPU and room sensors are {noisy and delayed}, and key variables are {unobserved}: {GPU heat generation}, {heat dissipation}, {local inlet-air effects}, and {time-to-throttle}. Actions, including {ambient setpoint}, {GPU frequency}, and {micro-batch size}, jointly affect heat, cooling, and latency. {ETCInfer therefore combines model-based prediction with belief-state control under partial observability, delay, and model error.}
We use POMDPs to {capture the partial observability. The contribution of ETCInfer lies in the ETCInfer state, action, constraint, and model design} for coupled cooling and LLM serving. {Compared with threshold control, deterministic optimization~\cite{boyd2004convex}, MPC~\cite{rawlings2017mpc}, robust MPC~\cite{bemporad1999robustmpc,mayne2005robustmpc}, and filtered MPC~\cite{kalman1960new}, ETCInfer retains prediction but learns belief states and a world model from observations, adapting online to stochastic token lengths, telemetry delay, airflow heterogeneity, DVFS variation, and hidden thermal dynamics.}
The system state ${\mathbf{s}_i}$ at time $i$ captures thermal and workload conditions as
\begin{equation}
\small
    \mathbf{s}_i=\Big[\ \{ \mathbf{\bar o}_i^x,  \mathbf{\tilde o}_i^x \}_{x\in\mathcal{G}_{\mathbf{j}}}, \bar \rho_{i},T_{\mathrm{set}},E_{1:i}(\mathbf{j},T_{\mathrm{set}}),\tilde P_{\mathrm{CRAC},i}(T_{\mathrm{set}}) \Big]
\end{equation}
where $\mathbf{\bar o}_i^x$ {is the sensor observation of GPU $x$ and includes sensed temperature used for safety checking},
$\bar \rho_{i}$ is the percentage of SLO left for this job,
$\mathbf{\tilde o}_i^x$ contains {GPU-related hidden quantities},
and $\tilde P_{\mathrm{CRAC},i}(T_{\mathrm{set}})$ is a room level quantity shared by all GPUs.
Typically, 
\begin{equation}
    \mathbf{\tilde o}_i^x = \{ \dot Q_G^x(i), \dot Q_D^x(i),  \tilde t_{thr}^x(i)\}
\end{equation}
where $\dot Q_G^x(i),\ \dot Q_D^x(i)$ are the heat generation/dissipation rate of GPU $x$ at time $i$. $\tilde t_{thr}^x(i)$ is the time to trigger GPU throttle.

Although those non-observable variants cannot be obtained directly, we could obtain them with physics-based estimation in the following section.

\section{Physics-informed modeling}
This section presents physics-based estimators that infer the {non-observable variables needed for control}. We integrate three layers, including GPU heat dynamics, facility cooling, and workload behavior based on physical rules. This {unified view} addresses what the model computes, how much heat the GPUs create, how much heat the room can dissipate, and how controls and scheduling trade performance against energy. 
Our model consists of five physics-informed components. 
\noindent\textbf{1) Heat Generation Model of GPU.} We estimate board heat from electrical power using a calibrated, frequency-driven die-power model, mapping die power to board power in a lightweight hardware-aware manner. 
\noindent\textbf{2) Heat Dissipation Model of Cooling.} We model the air-side path in a multi-GPU chassis, linking fan capacity and airflow geometry to maximum removable heat and describing how shared cooling is distributed across GPUs. 
\noindent\textbf{3) CRAC Power Model.} Room-level cooling power is expressed as a COP-based function of setpoint and IT heat with a clipped standby term. 
\noindent\textbf{4) Thermal Safety and Time to Throttle.} Estimated temperature and removed heat to obtain the predicted time before throttling, ensuring safe operation. 
\noindent\textbf{5) Latency Estimation Model of SLOs.} Prefill and decode latencies are combined into total latency and checked against SLOs, enabling evaluation of scheduling actions without executing the job.
The power model follows {frequency-driven GPU power modeling}{~\cite{hong2010integrated,leng2013gpuwattch,kandiah2021accelwattch,Guerreiro2019Modeling,Nabavinejad2022Coordinated}}, the cooling and CRAC terms follow {standard air-side heat-transfer and COP-based cooling approximations}{~\cite{moore2005making,pakbaznia2009minimizing,lin2023thermal,su2019research,ASHRAE_TC9_9_2021,EUCoC-DC-2024,wang2024greencooling35}}, and the latency model follows {roofline-style compute, memory, and communication accounting}{~\cite{williams2009roofline,yu2022orca,li2023vllm,nvidia_tensorrtllm_repo,zhong2024distserve,agrawal2024sarathi,Jiang2025Efficient}}. ETCInfer {calibrates their coefficients from telemetry} and uses the calibrated models for online scheduling.

\subsection{Thermal Estimation Models}
To estimate GPU thermal states, we account for two processes: heat produced by the GPUs and heat removed by the cooling system. First, we build a heat generation model that estimates a GPU’s heat output for a given hardware and workload in \S~\ref{subsec: heat_gen}. Then, we model heat dissipation through the onboard cooler to the ambient air. They allow us to estimate temperature over time.

\subsubsection{Frequency-driven Heat Generation Model of GPU}\label{subsec: heat_gen}
In modern GPUs, electrical energy powers transistors and memory, with losses as Joule heat in the silicon, VRMs, and the board. A GPU card has hundreds of components, including VRAM, the GPU die, fans, VRMs, inductors, and capacitors. Among these, the GPU die consumes most power, about 80--85\% in compute-heavy workloads, with VRAM at about 16\%. 
Therefore, thermal output in watts is approximately equal to the instantaneous electrical power draw of the GPU board as:
\begin{equation}
    \dot{Q}_G^x(i) \sim \tilde P_i^x = \tilde{P}_{die}^x(i)/\phi_{die}, 
\end{equation}
where $\phi_{die}=80\%$ is the proportion of the overall power from a single GPU die. 
We use a physics-based model{~\cite{hong2010integrated,leng2013gpuwattch,kandiah2021accelwattch,Guerreiro2019Modeling,Nabavinejad2022Coordinated}} that {estimates dynamic power as a calibrated function of operating frequency}. {After per-SKU profiling with GPU telemetry, this compact model tracks the operating-point-dependent power used by ETCInfer's scheduler.}
The physics-based approach to estimate the GPU die power is:
\begin{equation}
    \tilde{P}_{{die}}^x(i) = \beta_1^x f_i^x + \beta_2^x (f_i^x)^3 + \beta_3^x
\end{equation}
with $\beta_1^x,\beta_2^x$ obtained via calibration and $\beta_3^x$ representing leakage power. Voltage effects are captured in the fitted $\beta$.

\subsubsection{Heat Dissipation Model of Chassis Cooling}
\label{subsec:chassis_cooling_dissipation}

We model the air-side heat removal of a closed server chassis that contains multiple high-power GPUs, {following standard control-volume heat-transfer relations for datacenter cooling analysis~\cite{lin2023thermal,su2019research}}{ and datacenter thermal-guideline/placement assumptions~\cite{moore2005making,pakbaznia2009minimizing,ASHRAE_TC9_9_2021,EUCoC-DC-2024}}. The model connects fan specifications with the internal flow resistance of the chassis.
The goal of this model is to estimate the heat that can be carried away by the airstream per unit time, i.e., the heat dissipation rate.

\textbf{Assumptions.}
1) Fans operate at their maximum capability (fixed speed). 
2) No interaction between fans, where $N_{\mathrm{fan}}$ fans provide $N_{\mathrm{fan}}$ times the single-fan capacity. 
3) The chassis is a well-mixed control volume on the air side. 
4) A bypass factor $\gamma\in(0,1]$ accounts for leakage and recirculation.

\noindent\textit{{$N_{\mathrm{fan}}$ fans at maximum capability.}}
Let $u_{\mathrm{face}}$ be the measured or rated face velocity at maximum speed and $A_{\mathrm{fan}}$ the open flow area of the fan disc after hub and grill deductions. 
The total volumetric flow of $N_{\mathrm{fan}}$ {fans} is
\begin{equation}
   \dot V_N \;=\; N_{\mathrm{fan}}\, \dot V_{\max},\quad\text{where}\,   \dot V_{\max} \;=\; u_{\mathrm{face}}\,A_{\mathrm{fan}}.
  \label{eq:q_total}
\end{equation}

The corresponding mass flow is
\begin{equation}
  \dot m \;=\; \rho(\bar T^x_{\mathrm{in}})\,\dot V_N \;=\; \rho(\bar T^x_{\mathrm{in}})\,N_{\mathrm{fan}}\,\dot V_{\max}.
  \label{eq:m_dot}
\end{equation}
Therefore, the air-side capacity rate of each GPU could be computed on average as:
\begin{equation}\small
  \tilde{C}_{\mathrm{air}}^{tot} (i) \;=\; \gamma\,\dot m\,c_{air}(\bar T^x_{\mathrm{in}})  \quad [\mathrm{W/K}]
  \label{eq:c_air}
\end{equation}
where ${\rho(\bar T^x_{\mathrm{in}})}$ denotes air density as a function of inlet temperature and $c_{air}\approx1006\ \text{J kg}^{-1}\text{K}^{-1}$ for dry air near room temperature. And for each share of GPU over this chassis cooling system, the capacity rate is
\begin{equation}
    \tilde C^x_{\mathrm{air}}(i)=\tilde C_{\mathrm{air}}^{tot}(i)\,
\frac{\dot Q_G^x(i)}{\sum_y \dot Q_G^y(i)}.
\end{equation}
The airflow acts like a conveyor with capacity rate $\tilde C_{\mathrm{air}}(t) [\mathrm{W/K}]$.
Because heat must be generated before it can be removed, the heat dissipation rate is limited by both the generation and the air-side capacity, and is modeled as:
\begin{equation}
    \dot Q_{D}^x(i)=\min\{\dot Q^x_{G}(i),\tilde C^x_{\text{air}}(i)[\tilde T^x_{i}-\bar T^x_{\mathrm{in},i}]\}
\end{equation}
This bound enforces energy conservation and avoids unrealistically low outlet temperatures when cooling is saturated.

\subsubsection{GPU Throttle Trigger Time Estimation}
\label{subsec:throttle_time}

We adopt a lumped thermal model for GPU $x${, consistent with compact thermal modeling practice in datacenter thermal analysis~\cite{lin2023thermal}}. 
The temperature from time $i$ to the next time interval is estimated $i+1$ as:
\begin{equation}\label{eq:thermal}
    \tilde T^x_{i+1} = \tilde T^x_{i}+[\dot{Q}_G^{x}(i)-\dot{Q}_D^{x}(i)]/ C^x\cdot \Delta t ,
\end{equation}
Here, $\tilde T^x_{i+1}$ depends on the last-sampled temperature $\tilde T^x_i$ with measurement interval $\Delta t = t_{i+1}-t_{i}$, subject to the heat capacity of this substance, i.e., the temperature increase of this substance given a certain amount of heat. Here, $C^{x}=m^{x}c^{x}$ is the lumped thermal capacitance of the GPU, with $m^{x}$ the effective thermal mass and $c^{x}$ the specific heat capacity.

Let $T_{\mathrm{thr}}^{x}$ denote the thermal throttle threshold. The estimated time to reach the throttle threshold from $\tilde T^{x}_{i}$ is
\begin{equation}\label{eq:throttle_time}
\tilde t_{thr}^x(i) = \frac {T_{thr}^x - \tilde T^x_{i}}{\dot{Q}_G^{x}(i)-\dot{Q}_D^{x}(i)} \cdot C^x, \, \text{if } \dot{Q}_G^{x}(i)-\dot{Q}_D^{x}(i) \ge 0.
\end{equation}

\subsection{Power Model of CRAC}
\label{subsec:crac_simple}
We estimate the cooling device's electric power when the air temperature setpoint $T_{\mathrm{set}}$ maintains room inlet temperatures within a target range. {This model uses a COP-based approximation of datacenter cooling power with site-calibrated coefficients~\cite{lin2023thermal,wang2024greencooling35}.} {The setpoint and server-plus-cooling assumptions follow datacenter cost models and thermal guidelines~\cite{pakbaznia2009minimizing,moore2005making,ASHRAE_TC9_9_2021,EUCoC-DC-2024}.}

\textbf{Assumptions.}
1) The IT equipment generates a sensible total amount of heat rate $\dot Q_{\mathrm{IT}}(i)$ that the CRAC removes.
2) Latent load is negligible in the data hall.
3) Efficiency changes with $T_{\mathrm{set}}$ at a constant slope in a short range around $T_{\mathrm{ref}}$.

\noindent\textit{{a) Reference Coefficient of Performance: }}
$\mathrm{COP}_{\mathrm{ref}}$ is the ratio of heat removed to electric power at $T_{\mathrm{ref}}$.
A higher $\mathrm{COP}_{\mathrm{ref}}$ means lower power for the same load.
For example, a COP of 4 means the device removes 4 kW of heat by using 1 kW of electricity. COP rises when the temperature lift decreases, so increasing the evaporating or chilled-water temperature or lowering the condensing temperature typically improves COP.
 
\noindent\textit{{b) Temperature sensitivity: }}
Let $s$ denote the fractional change in cooling power per $^\circ\mathrm{C}$. In practice, $s \in [0.02, 0.05]$ and is calibrated from real data, as shown in Tab.~\ref{tab:CRAC_samples}.

\noindent\textit{{c) Model of CRAC Power: }}
The cooling system power at $t_i$ for setpoint $T_{\mathrm{set}}$ and heat removal $\dot Q_{IT}(i)$ is:
\begin{equation}
\label{eq:crac_power_linear}
\tilde P_{\mathrm{CRAC},i}(T_{\mathrm{set}})
=
\frac{\dot Q_{IT}(i)}{\mathrm{COP}_{\mathrm{ref}}}
\left[\,1 - s\,(T_{\mathrm{set}} - T_{\mathrm{ref}})\,\right].
\end{equation}
When raising the setpoint by $\Delta T>0$, the power reduces by approximately $s \times \Delta T$ in percent.
Lowering the setpoint increases power by the same rule.
Eq.\ref{eq:crac_power_linear} reflects the gain when the setpoint temperature rises.

\noindent\textit{d) Choice of the setpoint $T_{\mathrm{set}}$: }
Given an allowed inlet band $T_{\min} \le \bar T^x_{\mathrm{in}} \le T_{\max}$, choose $T_{\mathrm{set}}$ so that $\bar T^x_{\mathrm{in}} \le T_{\max}$.
For planning, evaluate \eqref{eq:crac_power_linear} at the selected $T_{\mathrm{set}}$.
For an upper bound on savings, we adopt $T_{\mathrm{set}}=T_{\max}$.

\begin{table}[t]
\caption{Typical CRAC samples}
\label{tab:CRAC_samples}
\resizebox{0.49\textwidth}{!}{
\begin{tabular}{lcc}
\hline
\textbf{System type} & $\mathrm{COP}_{\mathrm{ref}}$ & $s\ \bigl[\mathrm{per}\ ^\circ\mathrm{C}\bigr]$ \\
\hline
DX CRAC (air cooled) & $3.0$ to $4.0$ & $0.03$ to $0.04$ \\
Chiller + CRAH (air cooled) & $3.5$ to $5.5$ & $0.02$ to $0.03$ \\
Chiller + CRAH (water cooled) & $5.5$ to $7.0$ & $0.03$ to $0.05$ \\
High efficiency variable speed chiller & $7.0$ to $9.9$ & $0.04$ to $0.05$ \\
\hline
\end{tabular}
}

\end{table}

\subsection{Latency Estimation Model for Multi-GPU LLM Inference}
To satisfy SLO constraints, {estimating latency} $L_\mathbf{j}(\cdot)$ is necessary. We use a {roofline-style estimator} that separates {compute-bound, memory-bound, and communication-bound terms}, and then {calibrates residual runtime overheads from measured traces}{~\cite{williams2009roofline,li2023vllm,nvidia_tensorrtllm_repo}}{, following common LLM-serving latency, prefill/decode scheduling, and KV-cache considerations~\cite{yu2022orca,zhong2024distserve,agrawal2024sarathi,Jiang2025Efficient}}. Generally, LLM inference consists of two stages: \textit{prefill} and \textit{decoding}{, which recent serving systems schedule differently because of distinct batching, compute, and memory behavior~\cite{yu2022orca,zhong2024distserve,agrawal2024sarathi}}. 
1) In the prefill stage, the model processes the entire input, computes layer activations, and {builds the key--value (KV) cache for attention}. The {delay until the first output token appears} is called \textit{TTFT}.
2) In the decoding stage, the model generates tokens sequentially by reusing the KV cache from the prefill stage. Here, token latency is primarily determined by KV reads and increases with the effective context length. The time for each output token is called \textit{TPOT}.
In each stage, the time is the maximum of compute, memory, and any communication time due to parallelism.
Given per-GPU frequencies \( f_i^x \in [f_{\min}, f_{\max}] \) for all \( x \in \mathcal{G}_{\mathbf{j}} \) and \( i \in \mathcal{I} \), the effective peak FLOPs \( N_{FLOPS}^x(f^x) \) and peak VRAM bandwidth \( N_{BW}^x \) at the chosen precision can be determined. The relationship between frequency and peak FLOPs is profiled as:
\begin{equation}\small
 \label{eq:FLOPs}
    N_{FLOPS}^x \propto N_{core}^x \cdot f^x_i \cdot 2,
\end{equation}
where $N_{core}^x$ is the number of GPU core of GPU $x$, and {the factor 2 reflects fused multiply-add FLOP accounting in the profiled peak-FLOPs estimate}. Then, let
\begin{equation}
\Theta_{\mathrm{tot}}=\eta_c\sum_{x\in\mathcal{G}_{\mathbf{j}}} N_{FLOPS}^x(f^x),\quad\mathcal{B}_{\mathrm{tot}}=\eta_m\sum_{x\in\mathcal{G}_{\mathbf{j}}}N^x_{BW},
\end{equation}

where $\eta_c,\eta_m\in(0,1)$ capture kernel, utilization, and memory-efficiency losses. 

Let the model have ${N_L}$ layers, hidden size $d_{\text{model}}$, attention width $d_{\text{attn}}$ (often $d_{\text{attn}}=d_{\text{model}}$), and $N_{ne}$ non-embedding parameters. Each element uses $s_{\text{elt}}$ bytes (e.g., 2 for BF16/FP16, 1 for INT8). At decode step $k$, the context length is:
\begin{equation}
{N_{\mathrm{in},k}=N_{\mathrm{in}}+k-1},\qquad k=1,2, \dots,N_{\mathrm{out}}.
\end{equation}

\noindent\textit{a) Compute time: }
{Given} input prompt length $N_{in}$ and output prompt length $N_{out}$, the standard FLOPs accounting for a forward pass is:
\begin{align}\small
C_{\mathrm{pref}} (N_{in}) & = 2N_{ne}\,N_{\mathrm{in}} \;+\; 2{N_L}\,d_{\text{attn}}\,N_{\mathrm{in}}(N_{\mathrm{in}}+1), \label{eq:cprefill}\\\small
C_{\mathrm{dec}} (N_{out})&= \sum_{k=1}^{N_{\mathrm{out}}}\!\Bigl(2N_{ne} \;+\; 2{N_L}\,d_{\text{attn}}\,{N_{\mathrm{in},k}}\Bigr)
\label{eq:cdecode}
\end{align}
{We keep $C_{\mathrm{dec}}$ in summation form because each generated token has a different context length, while $C_{\mathrm{pref}}$ can be expanded directly over the fixed input prompt.}
The corresponding compute-bound times are estimated by
\begin{equation}
\tilde t_{\mathrm{cmp,pref}}=\frac{C_{\mathrm{pref}}}{\Theta_{\mathrm{tot}}},\qquad
\tilde t_{\mathrm{cmp,dec}}=\frac{C_{\mathrm{dec}}}{\Theta_{\mathrm{tot}}}.
\end{equation}

\noindent\textit{b) Memory time (KV cache traffic): }
During the prefill, we write KV for every token and layer with $N_{batch}$ batch size:
\begin{equation}
\text{Bytes}_{\mathrm{KV,prefill}}
= N_{batch}\cdot N_{\mathrm{in}}\cdot(2{N_L}\,d_{\text{model}})\,s_{\text{elt}}. \label{eq:kvprefill}
\end{equation}
During the decode step $k$, we read the cached K and V of all $N_{\mathrm{in},k}$ prior tokens and write the new token’s K and V:
\begin{equation}
\text{Bytes}_{\mathrm{KV,dec}}(k)
= N_{batch}\,(2N_L d_{\text{model}})\,s_{\text{elt}}\,(1+N_{\mathrm{in},k}).
\end{equation}
Then, when generating $N_{\mathrm{out}}$ tokens steps,
\begin{align}
\text{Bytes}_{\mathrm{KV,decode}}
&= \sum_{k=1}^{N_{\mathrm{out}}}\text{Bytes}_{\mathrm{KV,dec}}(k)
\label{eq:kvdecode}
\end{align}
The corresponding memory-bound times are
\begin{equation}
\tilde t_{\mathrm{mem,prefill}}=\frac{\text{Bytes}_{\mathrm{KV,prefill}}}{\mathcal{B}_{\mathrm{tot}}},
\tilde t_{\mathrm{mem,dec}}=\frac{\text{Bytes}_{\mathrm{KV,dec}}}{\mathcal{B}_{\mathrm{tot}}}.
\end{equation}

\noindent\textit{c) Communication time (multi-GPU): }
Let tensor-parallel degree be $p_{\mathrm{TP}}$ and pipeline-parallel stages be $p_{\mathrm{PP}}$ with $p_{\mathrm{TP}}p_{\mathrm{PP}}\le |\mathcal{G}_\mathbf{j}|$.

\noindent\textbf{Tensor parallel. }We abstract the intra-layer scheme that performs two all-reduces per layer as:
\begin{equation}
    \tilde t_{\mathrm{comm,TP}} \approx 2{N_L}\cdot\Bigl(\alpha_{\mathrm{TP}}\log p_{\mathrm{TP}}+\frac{\text{Bytes}_{\mathrm{act}}}{\beta_{\mathrm{TP}}}\cdot\frac{p_{\mathrm{TP}}-1}{p_{\mathrm{TP}}}\Bigr),
\end{equation}
where $\alpha_{\mathrm{TP}}$ and $\beta_{\mathrm{TP}}$ are the interconnect latency and bandwidth, and the per-layer activation size is
\begin{equation}
    \text{Bytes}_{\mathrm{act}} = S\,N_{batch}\,d_{\text{model}}\,s_{\text{elt}}/p_{\mathrm{TP}}
\end{equation}
where $S=N_{\mathrm{in}}$ for prefill and $S=N_{\mathrm{in},k}$ for decode step $k$.

\noindent\textbf{Pipeline parallel. }The bubble factor for $N_{mic}^x$ micro-batches:
\begin{equation}
\phi_{\mathrm{bubble}}=\frac{p_{\mathrm{PP}}-1}{N_{mic}^x+p_{\mathrm{PP}}-1},
\end{equation}
which multiplies the dominant per-stage time.
{Thus, increasing $N_{mic}^x$ reduces the pipeline-bubble penalty, but it also increases the work assigned to GPU $x$ in that step and can raise power, heat generation, and per-step latency.}
We collect all communication terms as $\tilde t_{\mathrm{comm,prefill}}$ and $\tilde t_{\mathrm{comm,dec}}$.

\noindent\textit{d) SLOs and total latency estimation: }
For each phase, we calculate the roofline maximum of compute and memory, and then add communication. With pipeline parallelism, multiply the dominant per-stage time by $(1+\phi_{\mathrm{bubble}})$.
\begin{align}\tiny
\tilde{t}_{\mathrm{prefill}} &= \max\{\tilde t_{\mathrm{cmp,prefill}},\,\tilde t_{\mathrm{mem,prefill}}\} + \tilde t_{\mathrm{comm,prefill}},\nonumber\\
\tilde{t}_{\mathrm{dec}}  &= \max\{\tilde t_{\mathrm{cmp,dec}},\,\tilde t_{\mathrm{mem,dec}}\} + \tilde t_{\mathrm{comm,dec}}.
\end{align}
The overall end-to-end latency could be estimated by:
\begin{equation}
\tilde{L}_{tot}=\tilde{t}_{\mathrm{prefill}}+\tilde{t}_{\mathrm{dec}}\,.
\end{equation}
The SLO metrics of a job can be estimated by:
\begin{equation}
\tilde{\mathrm{TTFT}}\;\approx\;\delta_0\;+\;\tilde t_{\mathrm{prefill}}, \,
\tilde{\mathrm{TPOT}}\;=\;\delta_1\;+\;\tilde t_{\mathrm{dec}}/N_{out},
\end{equation}
where $\delta_0$ and $\delta_1$ are {calibrated offsets} that represent {queueing, dispatch, first KV allocation, and other runtime overheads}.

\section{ETCInfer Solution}
\label{sec:solution}
\begin{figure}[t]
    \centering
     \includegraphics[width=0.975\linewidth]{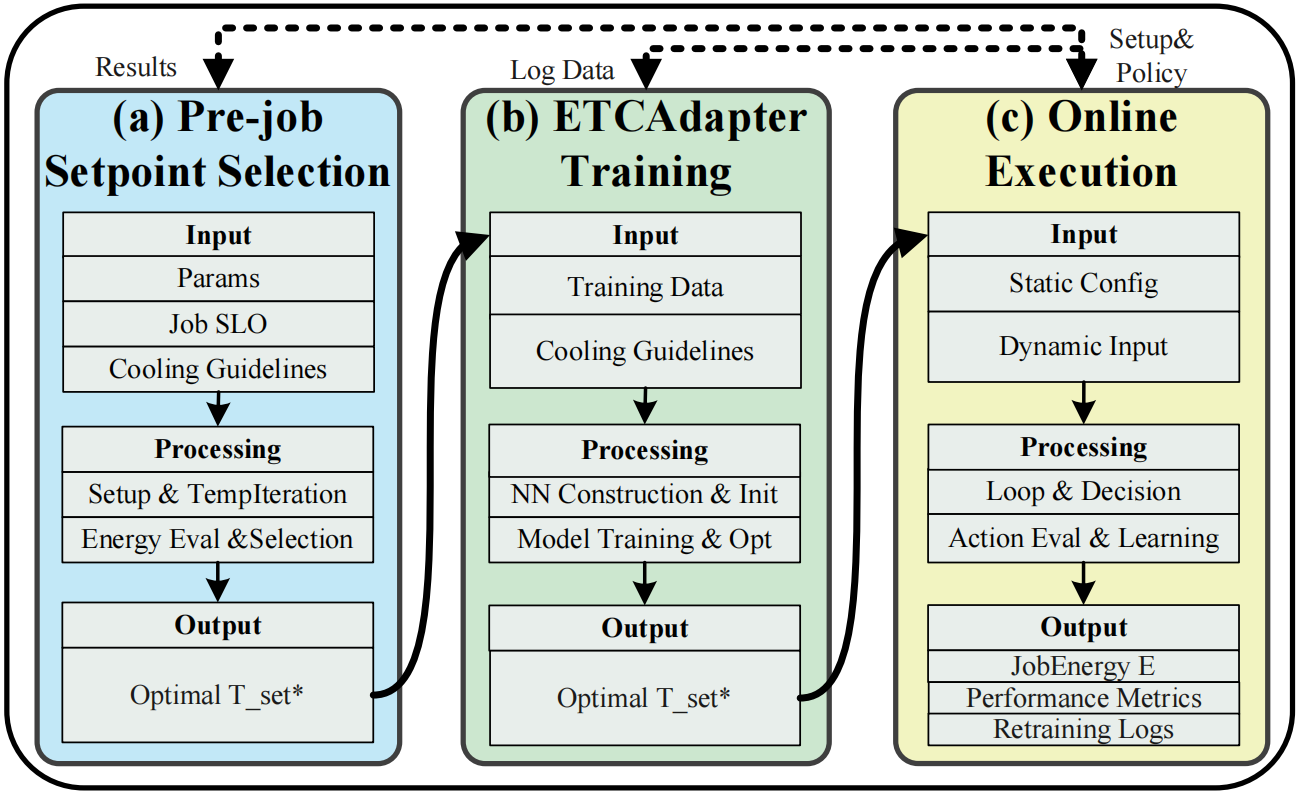}
    \caption{ETCInfer three-stage energy optimization.}
    \label{fig:etcinfer_tso}
    
\end{figure}

Our proposed solution, ETCInfer, comprises {three core stages} for energy optimization, detailed in Fig.~\ref{fig:etcinfer_tso}. The solution contains three parts. 
First, we select a {safe and promising temperature setpoint} before each job. 
Second, we train an online adaptation model, \textit{ETCAdapter}, that learns from logs and interactions with the simulator, informed by a learned world model. 
Third, we operate an online controller that adapts actions at every step using the {belief over hidden states}, while a safety layer enforces constraints.
The solution uses standard belief-state and model-based control ideas, but adapts them to a {joint control surface for LLM inference}: {ambient setpoint before execution}, {GPU frequency during execution}, and {micro-batch size at the serving layer}. This design lets ETCInfer evaluate {cooling energy, compute energy, SLO risk, and thermal safety} in one control loop.
{The setpoint stage handles slow facility response, the online frequency and micro-batch stage handles fast workload and thermal changes, and the safety layer prevents energy savings from violating latency or throttle constraints.}

\subsection{Pre-job Temperature Setpoint Selection}
\label{subsec:prejob}

Before all jobs start, we first build a {conservative starting policy $\pi_{0}$} that already meets the SLO target and the thermal limit. This starting policy is independent of ETCInfer designs, which typically serve as the default scheduling strategy in the existing scheduler and ensure that the system can run within SLO constraints, regardless of whether ETCInfer is active or not.
Next, we use it to construct a feasible set of setpoints and then select one that offers the potential for energy savings. 
Specifically, the starting policy $\pi_0$ fixes device frequency and micro-batch size for each GPU $x$ during a job. The starting policy $\pi_0$ records a safe and fixed frequency $f_{\mathrm{safe}}^x$ and balanced micro-batch size $N_{\mathrm{safe}}^x$ so that the latency satisfies $\mathrm{SLO}_{\mathbf{j}}$ and the hardware temperature never exceed $T_{\mathrm{thr}}^x$.

For each candidate $T\in[T_{\min},T_{\max}]$, iterate $\pi_0$  with~\eqref{eq:thermal}. 
Aggregate to the estimated job energy
\begin{equation}
\small
\tilde E(\mathbf{j},T|\pi_0) = \sum_{i=0}^{\tilde I_\mathbf{j}(\pi_0)} \Big ( w_{i}\,\tilde P_{\mathrm{CRAC},i-1}(T) + \sum_{x\in\mathcal{G}_{\mathbf{j}}} \hat P^{x}_{i-1} \Big ) \Delta t_{i},
\label{eq:pre_job_energy}
\end{equation}
and enforce the constraints $L_{\mathbf{j}}(\pi_0)\le \mathrm{SLO}_{\mathbf{j}}$ and $\hat T_i^x\le T_{\mathrm{thr}}^x$ under the thermal dynamics. The feasible set is $\mathcal{S}_T = \{T\in[T_{\min},T_{\max}] : \text{all constraints hold at risk level }\delta\}$. {Thus, the candidate optimization space is filtered by the current job, hardware state, latency budget, and predicted thermal headroom before a setpoint is selected.} We select
\begin{equation}
   T_{\mathrm{set}}^{\star}=\arg\min_{T\in\mathcal{S}_T}\ \tilde E(\mathbf{j},T\mid \pi_0). 
\end{equation}
After each job, the policy $\pi_0$ is {updated using the lowest-energy run}, logging hardware specs, $\mathrm{SLO}_{\mathbf{j}}$, and the setpoint $T_{\mathrm{set}}$, and yielding {updated $f_{\mathrm{safe}}^x$ and $N_{\mathrm{safe}}^x$}.

\subsection{Training and Designs of ETCAdapter}
\label{subsec:train}
\textit{ETCAdapter} is designed to learn an inner policy $\pi_{\mathrm{in}}$ that adapts at run time to reduce the true energy $E(\mathbf{j},T_{\mathrm{set}})$ while keeping $L_{\mathbf{j}}(\cdot)\le \mathrm{SLO}_{\mathbf{j}}$ and {$\bar T_i^x \le T_{\mathrm{thr}}^x$}.
{A belief $b_i$ summarizes uncertainty over $\mathbf{s}_i$} and is updated by a filter
\begin{equation}
b_{i+1}=\mathsf{Upd}(b_i,\mathbf{a}_i, \mathbf{\bar o}_{i+1}), \notag
\end{equation}
implemented as a Bayesian update.

\textbf{Learning process. }
We update $\pi_{\mathrm{in}}$ via {model-based RL in latent space}. The world model $g_{\theta}$ is trained on real and imagined data to capture thermal dynamics~\eqref{eq:thermal}. Short-horizon action sequences are simulated to compute cumulative rewards with penalties on latency $L_{\mathbf{j}}$ and predicted temperature $\hat T_i^x$. The policy is refined using advantage-weighted regression with a trust-region constraint, and {a barrier ensures $\bar T_i^x < T_{\mathrm{thr}}^x$ with high probability}.

\textbf{Model Module Designs}
Four neural modules implement the belief, the one-step predictor, the policy generator, and the critic. 

\subsubsection{Belief encoder $\phi$}
The belief encoder maps the history up to step $i$, denoted $\mathbf{h}_i=\{\mathbf{o}_{1\ :\ i},\,\mathbf{a}_{1\ :\ i-1},\,T_{\mathrm{set}}\}$, and outputs {a compact latent state $z_i=\phi(\mathbf{h}_i)$ that summarizes the posterior belief over hidden system variables}. This latent state conditions the one-step predictor $g_{\theta}$, the policy $\pi_{\mathrm{in}}$, and the critic $V$. 
The encoder is trained jointly with $g_{\theta}$ to retain information relevant to temperature, power, and control. Training minimizes a reconstruction and prediction loss on the next observation, temperatures, and powers, with a temporal smoothness regularizer on the latent dynamics:
\begin{align}
 \mathcal{L}_{\mathrm{enc}} & = \mathbb{E}\Big[ 
\underbrace{\|  {\mathbf{\hat o}}_{i+1}-\mathbf{o}_{i+1}\|_2^2}_{\text{optional observation}}
+ \underbrace{\|\hat T_{i+1}^{1\!:\!X}-T_{i+1}^{1\!:\!X}\|_2^2}_{\text{temperature}} \notag \\
& + \underbrace{\|\hat P_{i+1}^{1\!:\!X}-P_{i+1}^{1\!:\!X}\|_2^2}_{\text{power}}
\Big] + \alpha\,\mathbb{E}\big[\|z_{i+1}-z_i\|_2^2\big].
\end{align}
When raw observations are unavailable, the first term is omitted, leaving only prediction and smoothness losses, encouraging $z_i$ to be a Markovian summary for control.

\subsubsection{World model $g_{\theta}$ as a one-step predictor}
The world model $g_{\theta}$ inputs the current latent $z_i$, the control action $\mathbf{a}_i$, the setpoint $T_{\mathrm{set}}$, and the step size $\Delta t$. {It predicts and outputs the next latent $z_{i+1}$, the temperatures $\hat T^{1\!:\!X}_{i+1}$, and the GPU power $\hat P^{1\!:\!X}_{i+1}$.} When needed, it also produces the next observation $\hat{\mathbf{o}}_{i+1}$. The cooling power $P_{\mathrm{CRAC}}(T_{\mathrm{set}})$ is computed by a separate analytic model and is not a prediction of $g_{\theta}$.

The model is trained with a simple objective that matches its predictions to measured values and keeps them consistent with the thermal physics. The learning objective is
\begin{align}
\mathcal{L}_{\mathrm{wm}}
&=
\mathbb{E}\Big[
\|\hat T_{i+1}^{1\!:\!X}-T_{i+1}^{1\!:\!X}\|_2^2
+ \|\hat P_{i+1}^{1\!:\!X}-P_{i+1}^{1\!:\!X}\|_2^2 \nonumber \\ 
&+ \beta\,\|\hat{\mathbf{o}}_{i+1}-\mathbf{o}_{i+1}\|_2^2
\Big] + \lambda_{\mathrm{phys}}\,\mathbb{E}\Big[
\hat T_{i+1}^{1\!:\!X} - \tilde T_{i+1}^{1\!:\!X}
\Big]_2^{2}.
\end{align}
The first terms are standard prediction errors for temperature, power, and observations $\mathbf{o}$. The physics term enforces agreement with the thermal model, which improves robustness at unseen operating points. To quantify epistemic uncertainty, we also fit an ensemble $\{g_{\theta^{(k)}}\}_{k=1}^{K}$ on bootstrap splits. The variance across members guides risk-aware setpoint selection and provides signals for safety training.

\begin{algorithm2e}[t]
\caption{Online operation with energy- and latency-aware scoring}\label{algo:online}
\KwIn{$T_{\mathrm{set}}^{\star}$, policy $\pi_{\mathrm{in}}$, critic $V$, world model $g_{\theta}$, safety residual $\mathcal{S}$, thresholds $\{T_{\mathrm{thr}}^x\}$, job $\mathrm{SLO}_{\mathbf{j}}$.}
\KwOut{Realized job energy $E(\mathbf{j},T_{\mathrm{set}}^{\star})$, logs for future refresh.}
Fix $T_{\mathrm{set}}=T_{\mathrm{set}}^{\star}$ and start the job\;
\For{$i=0,1,\dots$ until the job ends}{
Measure $o_i$, update $b_i$, and compute $z_i=\phi(b_i)$\;
Generate action sequences guided by $\pi_{\mathrm{in}}$\;
Score each sequence with the return from $r(b_i,a_i;T_{\mathrm{set}})$ and the constraint costs based on $\tilde L_{tot}$ and {the predicted $T_i^x$}\;
Select the first action from the best safe sequence and apply the safety residual to obtain $\tilde a_i$\;
Execute $\tilde a_i$, log $(o_i,a_i)$, and update the model and critic on a small batch\;
\If{predicted latency threatens $\mathrm{SLO}_{\mathbf{j}}$} {shrink the batch size limits and/or increase the frequency within safe bounds, or lower the switching penalty.}
\If{{predicted or sensed temperature approaches $T_{\mathrm{thr}}^x$}} {{reduce frequency and/or shrink micro-batch limits to lower heat generation before throttling.}}
}
Compute $E(\mathbf{j},T_{\mathrm{set}}^{\star})$ by~\eqref{eq:job_energy} from the realized logs\;
\Return{$E(\mathbf{j},T_{\mathrm{set}}^{\star})$ and the new logs.}
\end{algorithm2e}

\subsubsection{Online Policy Generator $\pi_{\mathrm{in}}$}
The policy generator inputs the current latent state \( z_i \), the observable information \( \mathbf{\bar o}_i \), and the selected setpoint \( T_{\mathrm{set}} \), producing the control action \( \mathbf{a}_i \).
To align policy learning with the energy objective, we define the step reward as:
\begin{equation}
\small
    r = -( w_{i} \tilde P_{\mathrm{CRAC},i-1}(T_{\mathrm{set}}) + \sum_{x \in \mathcal{G}_{\mathbf{j}}} \tilde P^{x}_{i-1}) \Delta t - \lambda_{\mathrm{sw}} \Delta \mathbf{a}_i^2,
\end{equation}
where \( \Delta \mathbf{a}_i^2 = \|\mathbf{a}_i - \mathbf{a}_{i-1}\|_2^2 \) represents the change in control actions, and \( \lambda_{\mathrm{sw}} \) is the switching smoothness weight that controls how strongly the policy penalizes action changes. A larger \( \lambda_{\mathrm{sw}} \) ensures smoother frequency switches and batching updates.
Note that during inference, we always use the observed power $\bar P^{x}_{i-1}$.
The policy is trained with advantage-weighted regression under a trust region, plus a small behavior-cloning term that pulls it toward high-quality short-plan actions evaluated in $g_\theta$:
\begin{align}
\mathcal{L}_{\pi}
&= -\,\mathbb{E}\big[\,v_i \log \pi_{\mathrm{in}}(\mathbf{a}_i \mid z_i, \mathbf{o}_i, T_{\mathrm{set}})\big] \nonumber \\
& + \beta_{\mathrm{KL}}\, \mathrm{KL}\big(\pi_{\mathrm{in}} \,\|\, \pi_{\mathrm{ref}}\big) 
+ \lambda_{\mathrm{bc}}\, \mathbb{E}\big[ \| \mathbf{a}_i - \mathbf{a}_i^{\mathrm{plan}} \|_2^2 \big],
\end{align}
where $v_i = \exp(A_i / \tau)$ are critic-derived weights, $\pi_{\mathrm{ref}}$ is the previous policy for stable updates, and $\mathbf{a}_i^{\mathrm{plan}}$ are actions chosen by the planner within the learned world model.

\subsubsection{Critic $V$}
The critic takes as input the latent state $z_i$, the observable information $\mathbf{o}_i$, and the selected setpoint $T_{\mathrm{set}}$. Its role is to provide a scalar estimate $V(z_i)$ of the discounted return starting from step $i$. This value function supplies the advantage signal that is used to update the policy.
Training of the critic follows a temporal difference scheme with value expansion on imagined rollouts produced by the world model $g_{\theta}$. For a rollout horizon $H$, we construct the target
\begin{align}
G_i = \sum_{h=0}^{H-1} \gamma^{h} r_{i+h} + \gamma^{H} V(z_{i+H}),\,A_i = G_i - V(z_i).
\end{align}
and we optimize
\begin{align}
\mathcal{L}_{V} = \mathbb{E}\big[(V(z_i) - \mathrm{stop\_grad}(G_i))^2\big].
\end{align}
The critic employs short imagination inside $g_{\theta}$ so that the value targets are consistent with the learned dynamics and with the energy accounting. In the policy update, we use the advantages $A_i$ together with the weights $v_i$ obtained from the critic.

\subsection{Online end-to-end operation}
\label{subsec:online}

The goal of the online controller is to operate at the selected setpoint $T_{\mathrm{set}}^{\star}$ and at every step adapt the action according to the current belief, so that the energy is directly optimized and {all modeled latency and temperature constraints remain satisfied}.
At time $i$, the system receives the observation $\mathbf{o}_i$ and {updates the belief $b_i$ together with the latent representation $z_i = \phi(b_i)$}. Based on this latent state, the controller generates several candidate action sequences. Each candidate is evaluated by its short return $r(b_i, \mathbf{a}_i; T_{\mathrm{set}})$ and by the constraint costs that are obtained from the predicted latency $\hat L_b$ and the {predicted temperatures $T_i^x$} from~\eqref{eq:thermal}. {For the micro-batch component, candidate actions with larger $N_{mic}^x$ are preferred only when the predicted SLO slack and thermal headroom can absorb the extra per-step load. Otherwise, the controller shrinks $N_{mic}^x$ or raises frequency within safe bounds.} The controller selects the action that is safe and has the best score, executes this action on the system, and records the transition. The logged samples are then used to refresh the fast learners online.

\section{Implementation}
We implement ETCInfer through three components: {\textit{ETCInfer Scheduler}}, {\textit{ETCAdapter}}, and {\textit{Ambient Setpoint Control}}. Together, they connect serving control, learning-based adaptation, and ambient-temperature actuation.

\subsection{ETCInfer Scheduler Implementation}
ETCInfer uses common inference and cluster stacks: {vLLM for serving, Kubernetes for cluster management, KServe as the inference interface, and NVIDIA DCGM with a Kubelet extension to expose hardware status and frequency control}. {ETCInfer acts as an external coordination controller: it reads SLO metadata and telemetry, writes safe control decisions, and leaves native execution to the serving engine and cluster manager.}
{We released the ETCInfer source code.\footnote{\url{https://anonymous.4open.science/r/ETCInfer-2761/}}}

\noindent\textbf{LLM service layer.}
We deploy vLLM for high-throughput multi-GPU inference. {vLLM exposes a runtime control file that ETCInfer updates to select micro-batch size}. We extend the {KServe router to obtain SLO metadata} from requests, enabling ETCInfer to set SLO targets while KServe preserves routing fairness. ETCInfer runs as a {Kubernetes controller for Pod creation} without modifying core scheduling.

\noindent\textbf{Hardware access and control.}
We integrate {NVIDIA DCGM and a Kubelet plugin to collect per-GPU power, temperature, and utilization}. A sidecar reports telemetry every control interval. GPU frequency is exposed through {NVML functions}. The plugin restricts changes to safe ranges and restores baseline settings after jobs finish or risks appear.

\noindent\textbf{Execution flow.}
For each job, ETCInfer selects placement and an ambient setpoint, then KServe starts vLLM. During execution, ETCInfer performs {stepwise control}: it ingests telemetry, updates the physics model, and adjusts micro-batch size and GPU frequency while enforcing SLO and thermal constraints.

\subsection{ETCAdapter Implementation}
ETCAdapter is implemented in {Python using PyTorch} and runs separately from ETCInfer, communicating through lightweight RPC to avoid interfering with vLLM. Its encoder, latent updater, world model, policy, and critic are separate \texttt{torch.nn.Module}s built from MLP blocks with linear layers, PReLU, and layer normalization. A 128-dimensional CPU latent state is maintained. Device and rack embeddings are attention-pooled. A 20k-transition replay buffer supports micro-batch training.

We pretrain the encoder and world model on six hours of traces using Adam ($10^{-3}$, batch 256). Policy and critic use advantage-weighted regression ($5\times10^{-4}$, batch 128). Online, ETCAdapter updates only final layers with one to two gradient steps every two seconds.

\subsection{Ambient Setpoint Control Implementation}
We combine {CoolSIM CFD modeling with real cluster traces} to evaluate airflow, cold supply, hot return, and recirculation under LLM loads. ETCInfer selects a continuous inlet setpoint, and CFD computes the resulting temperature field. {The CFD solver is used for calibration and trace-driven ambient evaluation, not as a blocking component in the online vLLM request path.} We also validate behavior on a small server by adjusting chassis inlet airflow with PT100-sensor feedback.

\section{Evaluation and Simulation}\label{sec:eval}
\subsection{Evaluation Setup}\label{subsec:setup}
We evaluate ETCInfer using {real-trace CFD simulation and validation experiments}, covering energy, SLOs, and thermal safety across workloads and ambient conditions. The goal is to test whether ETCInfer improves efficiency under realistic facility and serving-stack constraints.

\begin{table}[t]
\centering
\caption{{Workload profiles and service-level-objective thresholds}}
\label{tab:workloads_slo}
\resizebox{\linewidth}{!}{
\setlength{\tabcolsep}{0.5mm}{
\begin{tabular}{cccc}
\toprule
\textbf{Inference Task } & \textbf{Chat Dialog }       & \textbf{Code Assist}              & \textbf{Summarization} \\ \midrule
\textbf{Trace}           & \textit{OASST1}             & \textit{PromptSet}                & \textit{WebGPT}                \\
\textbf{Arrival Pattern} & Poisson            & Bursty with spikes       & Poisson     \\
\textbf{SLO (TTFT) }     & $\leq$800 ms & $\leq$1500 ms      & $\leq$2000 ms  \\
\textbf{SLO ({TPOT})}      & $\leq$50 ms  & $\leq$60 ms        & $\leq$80 ms    \\
\textbf{SLO ({E2E})}     & $\leq$4 s    & $\leq$8 s          & $\leq$12 s    \\ 
\bottomrule
\end{tabular}
}}

\end{table}

\noindent\textbf{Validation testbeds.}
We validate ETCInfer on a {small-scale physical workstation} with an Intel Core i9-13900K CPU, 128 GB RAM, four RTX3090 GPUs, and four RTX4090 GPUs ($V1$, $V2$), each with 24 GB VRAM. Models are quantized to fit memory. The tower chassis has front intake, rear exhaust, and 120 mm fans inside a $2\,\mathrm{m}\times2\,\mathrm{m}$ enclosure. ETCInfer selects continuous inlet setpoints from $18^\circ\text{C}$ to $48^\circ\text{C}$.

\noindent\textbf{CFD simulator configuration.}
To capture {airflow and recirculation}, we model two $2\,\mathrm{m}\times2\,\mathrm{m}\times2.4\,\mathrm{m}$ rooms, $R1$ and $R2$. $R1$ has strong containment and uniform inlet temperatures, while $R2$ has partial containment, stronger recirculation, and higher inlet temperatures at the same setpoint. Each chassis includes 8 NVIDIA H100-class GPU heat sources and a 350 W Intel Xeon 8480 CPU. GPU heat follows measured board-power traces scaled to the H100 SXM power class~\cite{nvidia_h100_pb11133}. We use {open LLM prompt traces and a production cluster trace}: OASST1~\cite{kopf2023openassistant}, PromptSet~\cite{pister2024promptset}, WebGPT~\cite{nakano2021webgpt}, and Alibaba2020~\cite{alibaba2020gpu}. We derive arrivals, parameterize power/temperature dynamics, and replay setpoints from $18^\circ$C to $47^\circ$C.

\noindent\textbf{LLM inference models.}
{We use Llama-3.1-Instruct~\cite{grattafiori2024llama3} for Chat Dialog (\textit{\textbf{CD}}) and search summarization (\textit{\textbf{SUM}}), and Code-Llama-Instruct~\cite{roziere2023codellama} for code assist (\textit{\textbf{CA}}).} These traces cover distinct token statistics, arrival patterns, and workload-specific SLOs in Table~\ref{tab:workloads_slo}.

\begin{figure*}[t]
    \centering
    \subfigure[R1]{\includegraphics[width=0.245\linewidth]{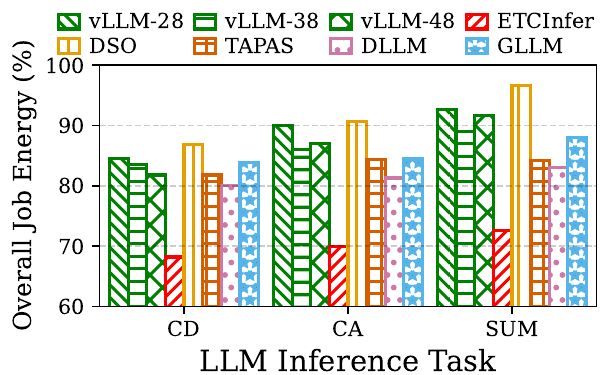}}
    \subfigure[R2]{\includegraphics[width=0.245\linewidth]{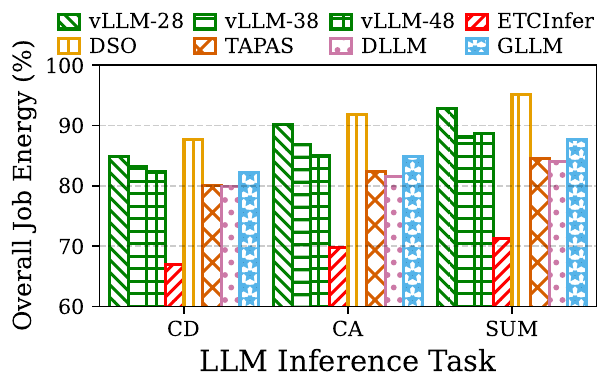}}
    \subfigure[V1]{\includegraphics[width=0.245\linewidth]{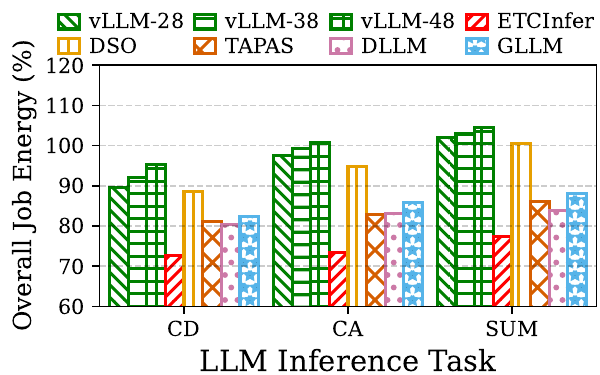}}
    \subfigure[V2]{\includegraphics[width=0.245\linewidth]{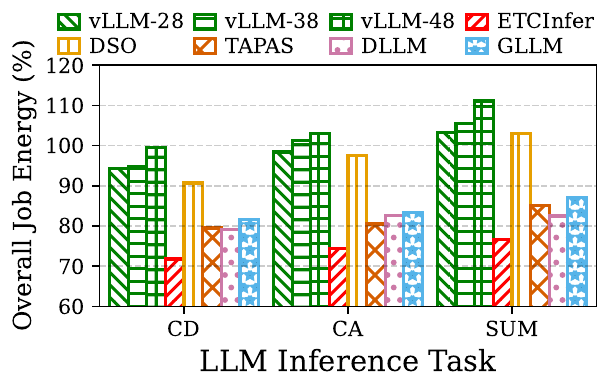}}
    
    \caption{Comparison of overall job energy consumption across four scenarios (R1, R2, V1, V2) and three tasks (CD, CA, SUM).}
    \label{fig:overall_eng}
    
\end{figure*}

\begin{figure*}[t]
    \centering
    \subfigure[R1]{\includegraphics[width=0.245\linewidth]{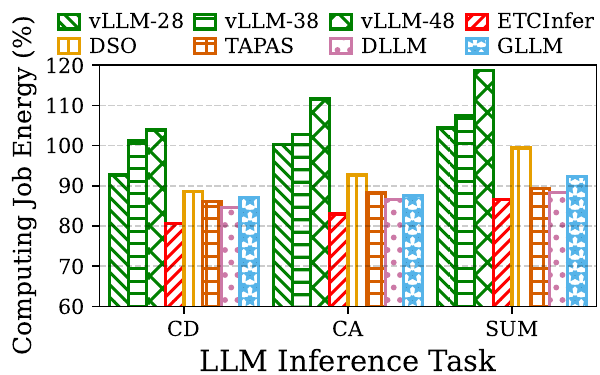}}
    \subfigure[R2]{\includegraphics[width=0.245\linewidth]{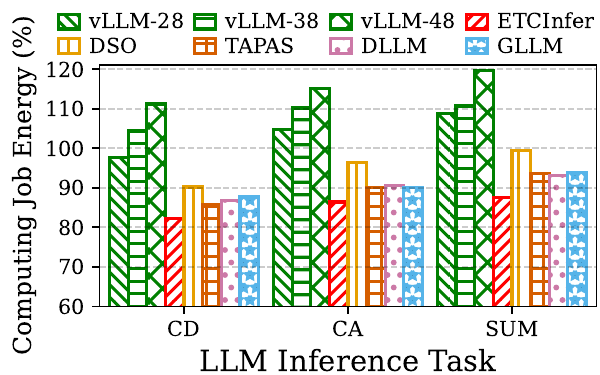}}
    \subfigure[V1]{\includegraphics[width=0.245\linewidth]{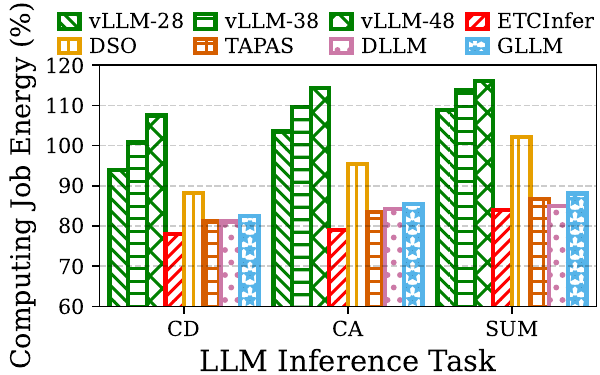}}
    \subfigure[V2]{\includegraphics[width=0.245\linewidth]{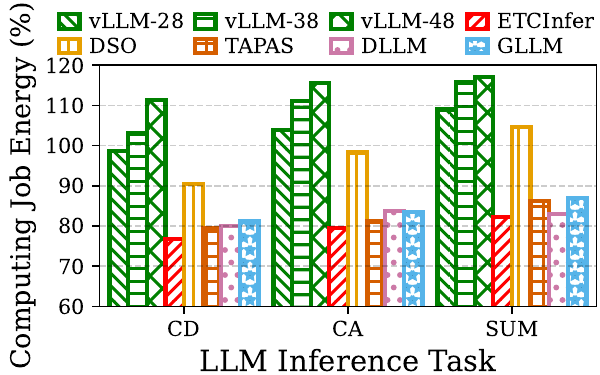}}
    
    \caption{Comparison of computing energy consumption across four scenarios (R1, R2, V1, V2) and three tasks (CD, CA, SUM).}
    \label{fig:comp_eng}
    
\end{figure*}

\begin{figure*}[t]
    \centering
    \subfigure[R1]{\includegraphics[width=0.245\linewidth]{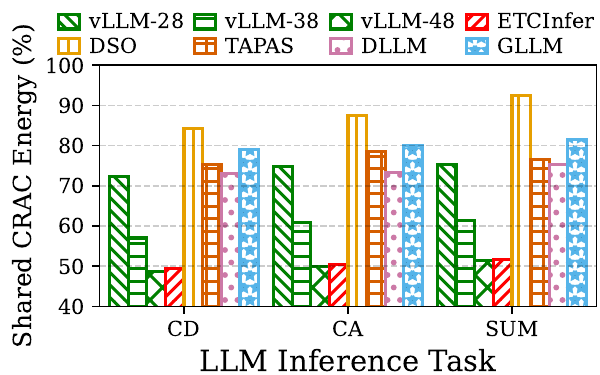}}
    \subfigure[R2]{\includegraphics[width=0.245\linewidth]{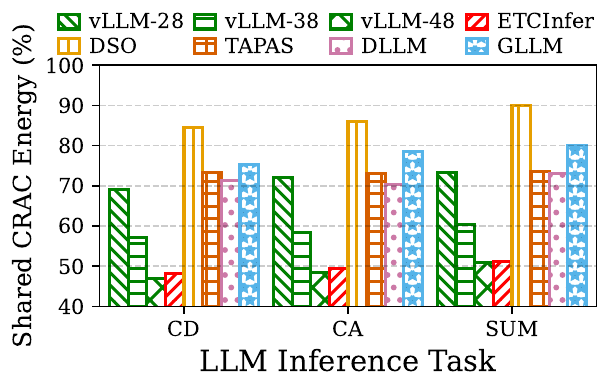}}
    \subfigure[V1]{\includegraphics[width=0.245\linewidth]{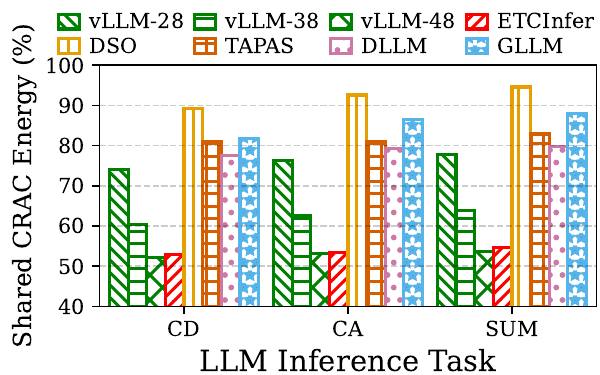}}
    \subfigure[V2]{\includegraphics[width=0.245\linewidth]{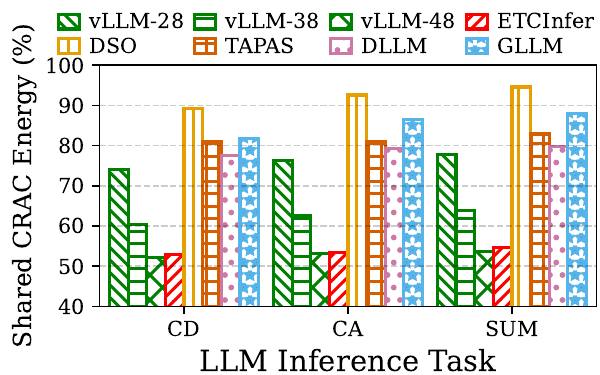}}
    
    \caption{Comparison of shared CRAC energy consumption across four scenarios (R1, R2, V1, V2) and three tasks (CD, CA, SUM).}
    \label{fig:crac_eng}
    
\end{figure*}

\noindent\textbf{Metrics.}
In our evaluation, we measure energy from three views: average job overall energy, average job computing energy, and average CRAC energy. All values are with respect to (wrt) the default vLLM setting under an $18^\circ\text{C}$ ambient setpoint.
{For request $r$ in workload class $\mathbf{j}$, let $a_r$ be the arrival time at the inference frontend, $\tau_{r,k}$ be the emission time of the $k$-th output token, and $N^{\mathrm{out}}_r$ be the number of generated tokens. We define the latency metrics as}
\begin{align}
{\mathrm{TTFT}_r} &{= \tau_{r,1}-a_r,} \notag\\
{\mathrm{TPOT}_r} &{= \frac{\tau_{r,N^{\mathrm{out}}_r}-\tau_{r,1}}{N^{\mathrm{out}}_r-1},\quad N^{\mathrm{out}}_r>1,} \notag\\
{L^{\mathrm{e2e}}_r} &{= \tau_{r,N^{\mathrm{out}}_r}-a_r.} \notag
\end{align}
{The workload class $\mathbf{j}$ specifies thresholds $(\theta^{\mathrm{TTFT}}_{\mathbf{j}},\theta^{\mathrm{TPOT}}_{\mathbf{j}},\theta^{\mathrm{e2e}}_{\mathbf{j}})$, as listed in Table~\ref{tab:workloads_slo}. Request $r$ satisfies the latency SLO only if $\mathrm{TTFT}_r\le\theta^{\mathrm{TTFT}}_{\mathbf{j}}$, $\mathrm{TPOT}_r\le\theta^{\mathrm{TPOT}}_{\mathbf{j}}$, and $L^{\mathrm{e2e}}_r\le\theta^{\mathrm{e2e}}_{\mathbf{j}}$.} We measure serving performance through {metric-specific SLO violation rates for TTFT, TPOT, and end-to-end latency. For a metric $m$, the violation rate is the fraction of requests whose measured value exceeds the corresponding threshold $\theta^m_{\mathbf{j}}$}. Thermal safety is reported as throttle exposure time.

\noindent\textbf{Baselines.}
{We compare ETCInfer with recent LLM serving systems and with commonly used thermal and energy control strategies.} Those baselines without specification would execute in the default temperature of $28^\circ\text{C}$.

\noindent $~\bullet~$\textbf{vLLM Default} is a throughput-first serving baseline that uses the original vLLM framework running under fixed ambient conditions {$\{18^\circ\text{C},28^\circ\text{C},\,38^\circ\text{C},\,48^\circ\text{C}\}$, denoted as vllm-18/28/38/48}. It serves as the baseline with stable environment settings where cooling is independent from GPU scheduling.

\noindent $~\bullet~$\textbf{DSO}~\cite{wang2024dso} is a GPU energy-efficiency optimizer that fuses static program information with runtime signals for DVFS decisions. In our comparison, it controls device-side frequency but does not coordinate with ambient setpoints or airflow models.

\noindent $~\bullet~$\textbf{TAPAS} \cite{stojkovic2025tapas} is a thermal- and power-aware scheduler. It predicts power and device temperature and allocates requests to a cooler rack to avoid thermal violations within SLOs. It focuses on device safety and power efficiency, but does not adjust ambient conditions or use airflow models.

\noindent $~\bullet~$\textbf{DLLM}  \cite{stojkovic2025dynamollm} is a cluster energy control approach using elastic reconfiguration and GPU frequency selection. The method adapts GPU assignment and DVFS to lower energy use while preserving service quality. Its control stays inside the cluster and does not incorporate facility-side thermal management.

\noindent $~\bullet~$\textbf{GLLM} \cite{tian2024greenllm} is an energy-aware pruning method for LLMs. We include it as a model-side efficiency baseline that lowers compute demand, but it does not perform runtime placement, GPU-frequency control, or ambient setpoint coordination.

\subsection{Evaluation Results}

\subsubsection{Improvement of Energy Efficiency}
We compare ETCInfer with {existing baselines across R1, R2, V1, and V2}. Energy is normalized to vLLM-18 in each scenario, so Fig.~\ref{fig:overall_eng}--\ref{fig:crac_eng} report relative job energy.

In R1 and R2, cooling is a large share of total energy. Higher-ambient vLLM settings reduce CRAC power but increase compute energy because jobs run longer. Thus, total savings are limited, and long traces such as SUM can even consume more energy; the best vLLM saving is 18.2\%. DSO, TAPAS, DLLM, and GLLM slightly reduce compute energy but leave cooling mostly unchanged, limiting gains to about 12.0\%--20.1\%. {ETCInfer achieves the largest savings in R1 and R2 by coordinating device-side and facility-side actions.} It lowers GPU power in non-critical phases, cutting compute energy by at least 12.5\%, and raises ambient setpoints safely, reducing cooling energy by 49.9\%. Together, these effects save up to 33.1\% total energy in R2. In V1 and V2, GPU power dominates, so ambient-only control is weaker and vLLM-28/48 can exceed vLLM-18. ETCInfer still improves compute efficiency and preserves cooling savings.

\begin{figure*}[t]
    \centering
    \subfigure[R1]{\includegraphics[width=0.245\linewidth]{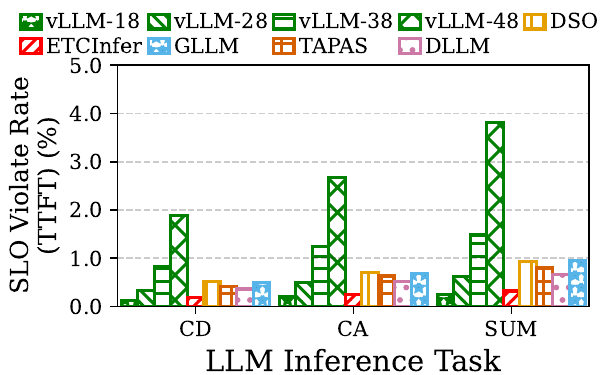}}
    \subfigure[R2]{\includegraphics[width=0.245\linewidth]{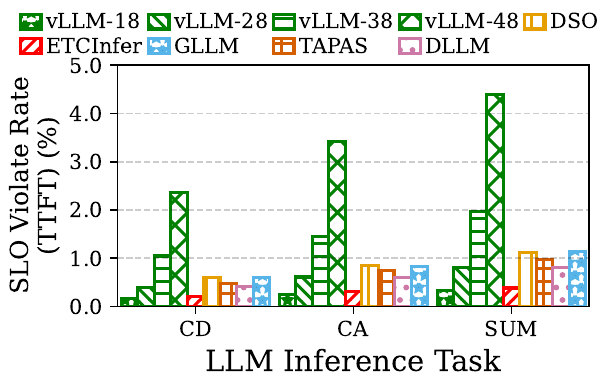}}
    \subfigure[V1]{\includegraphics[width=0.245\linewidth]{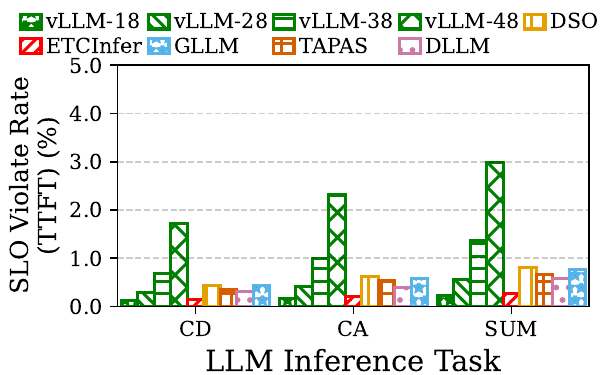}}
    \subfigure[V2]{\includegraphics[width=0.245\linewidth]{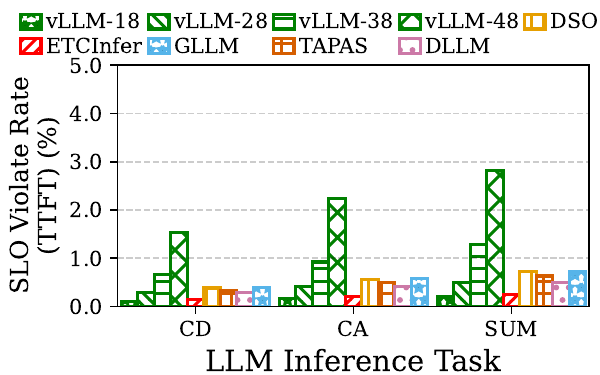}}
    
    \caption{Comparison of SLO violation rates for TTFT across four scenarios (R1, R2, V1, V2) and three tasks (CD, CA, SUM).}
    \label{fig:slo_ttft}
    
\end{figure*}

\begin{figure*}[t]
    \centering
    \subfigure[R1]{\includegraphics[width=0.245\linewidth]{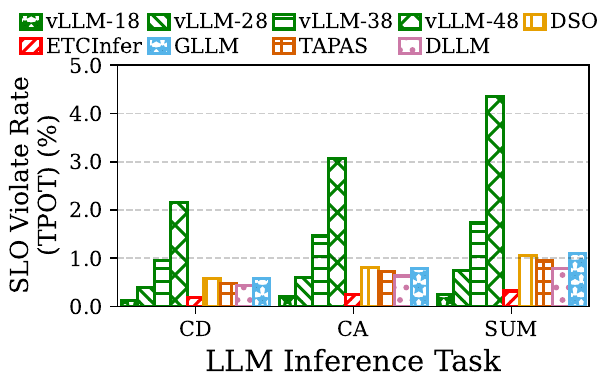}}
    \subfigure[R2]{\includegraphics[width=0.245\linewidth]{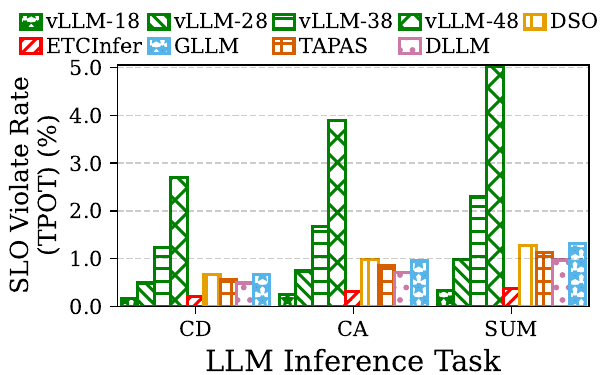}}
    \subfigure[V1]{\includegraphics[width=0.245\linewidth]{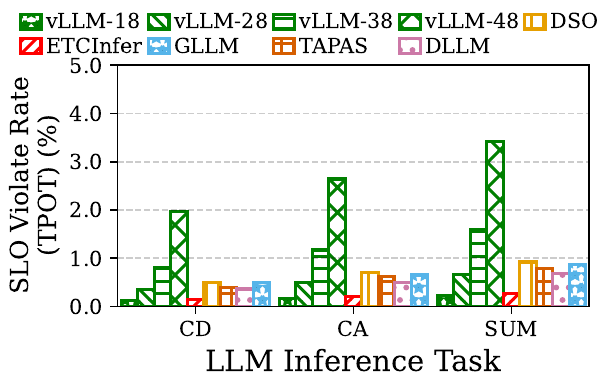}}
    \subfigure[V2]{\includegraphics[width=0.245\linewidth]{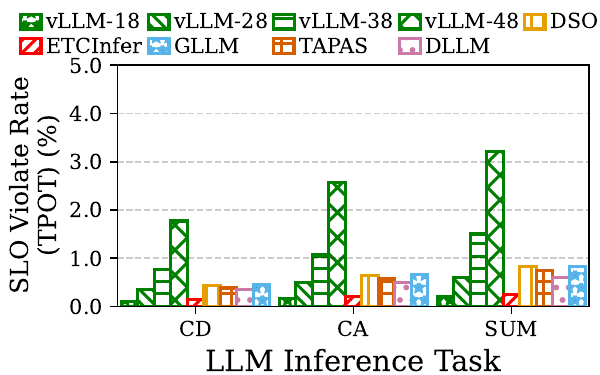}}
    
    \caption{Comparison of SLO violation rates for TPOT across four scenarios (R1, R2, V1, V2) and three tasks (CD, CA, SUM).}
    \label{fig:slo_tpot}
    
\end{figure*}

\begin{figure*}[t]
    \centering
    \subfigure[R1]{\includegraphics[width=0.245\linewidth]{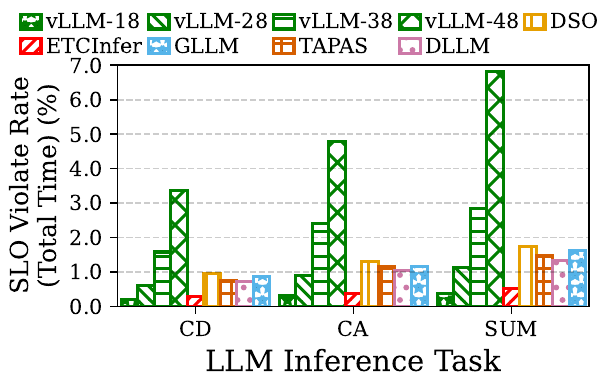}}
    \subfigure[R2]{\includegraphics[width=0.245\linewidth]{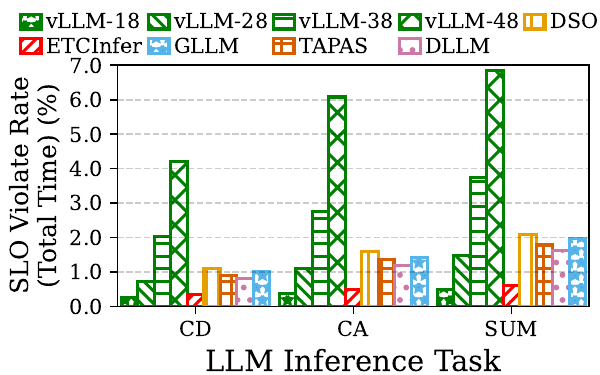}}
    \subfigure[V1]{\includegraphics[width=0.245\linewidth]{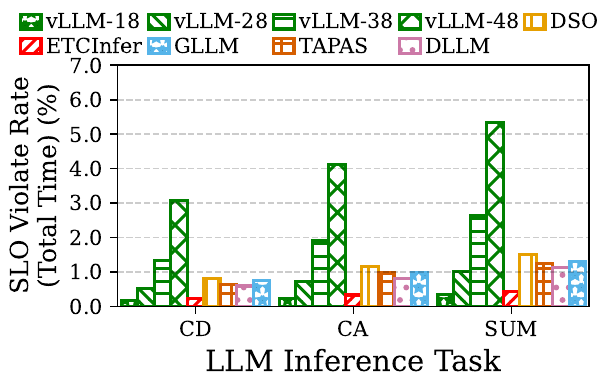}}
    \subfigure[V2]{\includegraphics[width=0.245\linewidth]{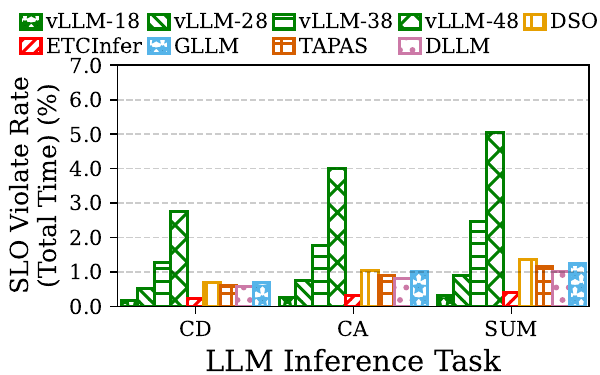}}
    
    \caption{Comparison of SLO violation rates for overall end-to-end latency across four scenarios (R1, R2, V1, V2) and three tasks (CD, CA, SUM).}
    \label{fig:slo_tt}
    
\end{figure*}

\begin{figure*}[t]
    \centering
    \subfigure[R1]{\includegraphics[width=0.245\linewidth]{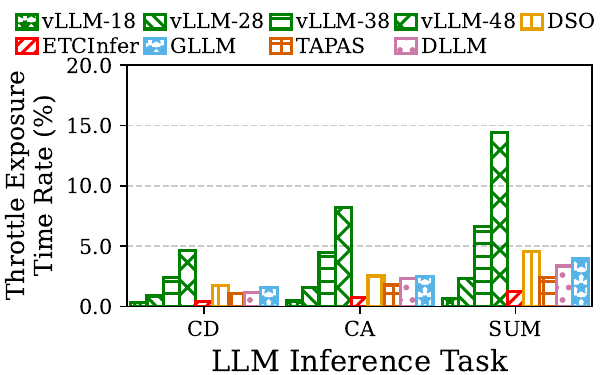}}
    \subfigure[R2]{\includegraphics[width=0.245\linewidth]{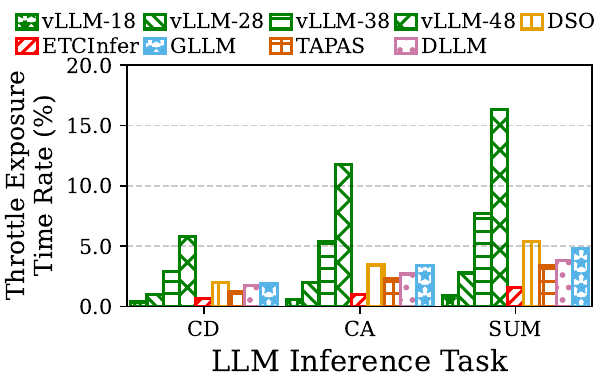}}
    \subfigure[V1]{\includegraphics[width=0.245\linewidth]{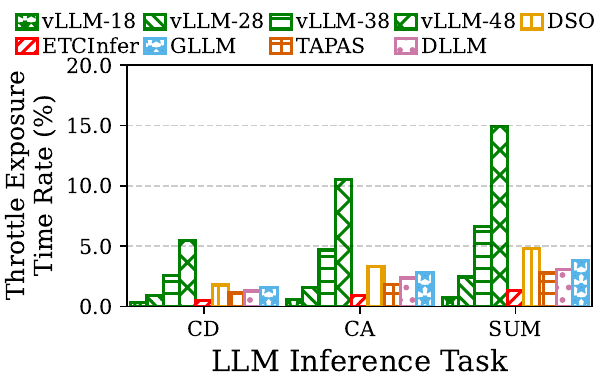}}
    \subfigure[V2]{\includegraphics[width=0.245\linewidth]{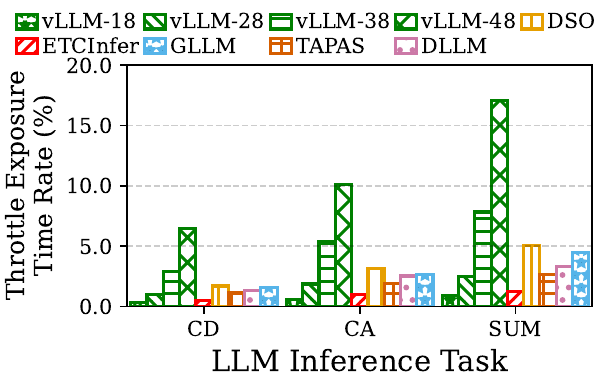}}
    
    \caption{Comparison of throttle exposure time across four scenarios (R1, R2, V1, V2) and three tasks (CD, CA, SUM).}
    \label{fig:throttle_expt}
    
\end{figure*}

\subsubsection{Improvement of SLO violation rate}
We evaluate TTFT, TPOT, and end-to-end SLO violations in Fig.~\ref{fig:slo_ttft}--\ref{fig:slo_tt}; lower is better. vLLM-18 keeps violations low, with 0.3\% total violations on CD and at most 0.5\% on long SUM traces. Raising ambient without thermal awareness quickly worsens latency. In R1 and R2, vLLM-28/38 often reach 0.6\%--3.8\% violations, while vLLM-48 exceeds 7.9\% in the worst R2 case. The effect is stronger in V1 and V2, where long requests already push GPUs near compute and thermal limits. DSO reduces power but slows decoding, causing about $3\times$ the vLLM-18 violation rate in R2 and V2. TAPAS, DLLM, and GLLM improve over vLLM-38/48, usually staying within 0.6\%--2.0\%, but do not recover vLLM-18 quality. {ETCInfer keeps violations close to vLLM-18 in all scenarios}: totals stay within 0.2\% of vLLM-18 in R1/R2 and below 0.7\% at the highest setpoint.

\subsubsection{Improvement of Hardware Thermal Safety}
We measure thermal safety by throttle exposure time in Fig.~\ref{fig:throttle_expt}, i.e., the fraction of execution above a high-temperature threshold. Higher exposure indicates sustained thermal stress and greater aging or shutdown risk. Higher-ambient vLLM settings sharply increase exposure, especially in R2 and V2 and on the long SUM trace. DSO suppresses spikes but extends runtime, so heat still accumulates. TAPAS, DLLM, and GLLM reduce exposure but remain riskier than ETCInfer. In the worst case, V2 with vLLM-48 on SUM spends 17.1\% of execution above the threshold, while ETCInfer reduces this to 1.2\%, a 92.9\% drop. {ETCInfer consistently achieves the lowest throttle exposure below 1.6\% in R2 and V2.}

\begin{figure*}[t]

\begin{minipage}[t]{0.49\textwidth}
    \centering
    \subfigure[Overall Energy]{\includegraphics[width=0.32\linewidth]{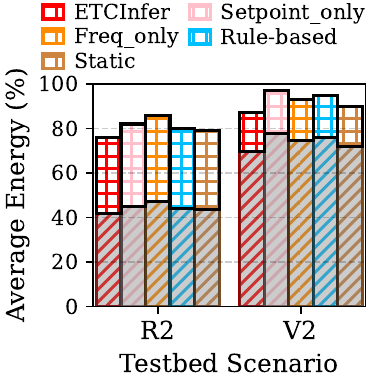}}
    \subfigure[SLO Violation Rate]{\includegraphics[width=0.32\linewidth]{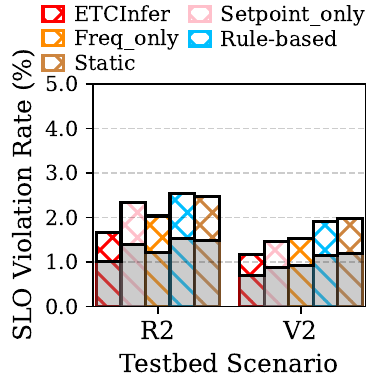}}
    \subfigure[Throttle Expose Time]{\includegraphics[width=0.325\linewidth]{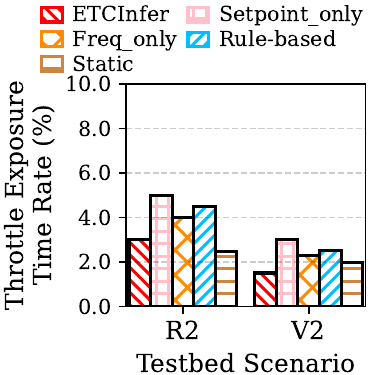}}
    \caption{Effects of {joint control}.}
    \label{fig:aba_joint}
\end{minipage}
\begin{minipage}[t]{0.49\textwidth}
    \centering
    \subfigure[Overall Energy]{\includegraphics[width=0.32\linewidth]{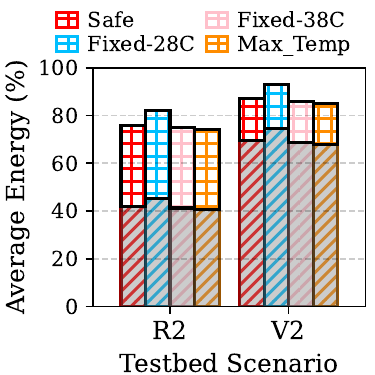}}
    \subfigure[SLO Violation Rate]{\includegraphics[width=0.32\linewidth]{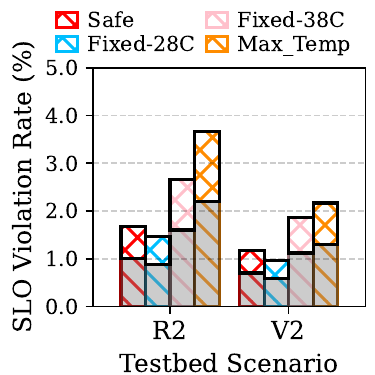}}
    \subfigure[Throttle Expose Time]{\includegraphics[width=0.325\linewidth]{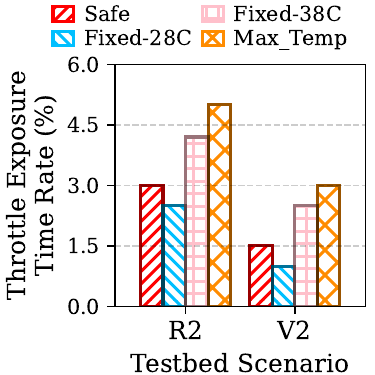}}
    \caption{Effects of the ambient setpoint setup.}
    \label{fig:aba_sopa}
\end{minipage}

\end{figure*}

\begin{figure}[t]
    \centering
    \subfigure[Overall Energy]{\includegraphics[width=0.32\linewidth]{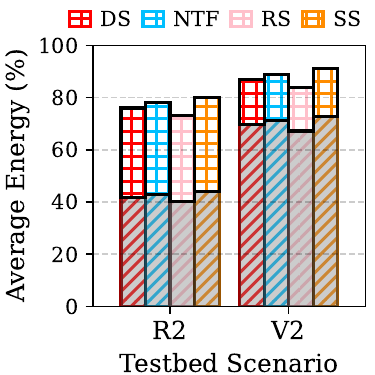}}
    \subfigure[SLO Violation Rate]{\includegraphics[width=0.32\linewidth]{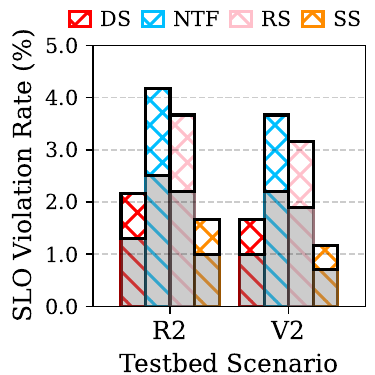}}
    \subfigure[Throttle Expose Time]{\includegraphics[width=0.325\linewidth]{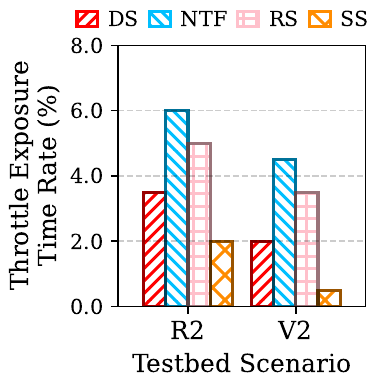}}
    
    \caption{Effects of SLO and thermal-safety knobs.}
    \label{fig:aba_knobs}
    
\end{figure}

\begin{table}[t]
\centering
\caption{{Validation comparison with model-based control alternatives.}}
\label{tab:control_alternatives}
\resizebox{\linewidth}{!}{
\begin{tabular}{lccc}
\toprule
\textbf{{Control method}} &
\textbf{{Avg. energy saving}} &
\textbf{{Avg. SLO violation}} &
\textbf{{Avg. throttle-exposure reduction}} \\
\midrule
{Greedy threshold} & {18.6\%} & {1.84\%} & {54.3\%} \\
{Deterministic optimization} & {22.1\%} & {1.26\%} & {63.5\%} \\
{MPC} & {25.4\%} & {0.92\%} & {72.8\%} \\
{Robust MPC+filter} & {27.6\%} & {0.74\%} & {78.9\%} \\
{ETCInfer} & {31.8\%} & {0.48\%} & {89.7\%} \\
\bottomrule
\end{tabular}}

\end{table}

\subsection{Ablation Study}
We analyze ETCInfer components to {measure the effect of each design choice}. All ablations use R2 and V2, averaged over three traces.

\footnotetext{In (a), gray stripes denote computing energy (bottom) and CRAC energy (top); in (b), they denote TTFT and TPOT. This notation also applies to Figs. 15--17.}

\subsubsection{Effectiveness of Joint and Adaptive Control}
We evaluate {joint ambient-setpoint, workload-scheduling, and online-adaptation control} in Fig.~\ref{fig:aba_joint}. ETCInfer is compared with \textit{Setpoint\_only}, \textit{Freq\_only}, \textit{Rule-based}, and \textit{Static}. Full ETCInfer has the lowest total energy by reducing compute and CRAC costs. In R2, Setpoint\_only leaves energy 7.9\% higher and increases SLO violations by 40.0\%. In V2, Freq\_only is 6.9\% worse and increases throttling. Rule-based performs between single-actuator baselines and Static, while Static still uses 3.7\% more energy, showing the online adaptation benefits.

\subsubsection{Impact of Ambient Setpoint Setup}
We compare four pre-job setpoint policies in R2 and V2: \textit{Safe}, fixed 28$^\circ$C, fixed 38$^\circ$C, and \textit{Max\_Temp}. Fig.~\ref{fig:aba_sopa} reports total, compute, and CRAC energy, SLO violations, and throttling. \textit{Safe} nearly minimizes energy while keeping violations and throttling low. Max\_Temp saves only 2.6\% and 2.3\% total energy in R2 and V2, but raises SLO violations by 1.2$\times$ and throttle exposure by over 65.7\%. Fixed-28C improves reliability, reducing SLO violations by 12.0\% and throttling by 16.7\% in R2, but increases total and CRAC energy by 7.9\% and 6.9\%. Fixed-38C saves only 1.1--1.3\% energy while raising violations by 59.8\% and throttling by 40.5--63.7\%. Thus, fixed setpoints cannot balance energy and reliability.

\subsubsection{Sensitivity to SLO and Thermal Safety Knobs}
We compare \textit{default\_safety (DS)}, \textit{strict\_safety (SS)}, \textit{relaxed\_safety (RS)}, and \textit{no\_ttt\_feature (NTF)} in Fig.~\ref{fig:aba_knobs}. SS tightens SLO budgets by 10\% and lowers the thermal limit by $3^\circ$C; RS expands SLOs by 20\% and raises it by $3^\circ$C. DS balances energy, violations, and throttling. SS uses 4.9\% more energy but reduces violations by 26.5\%. RS saves 3.7\% energy but increases violations and near-limit time. NTF worsens all metrics, confirming the importance of time-to-throttle features.

\begin{table}[t]
\centering
\caption{{Latency-estimator prediction errors on workload traces.}}
\label{tab:latency_error}
\resizebox{0.8\linewidth}{!}{
\begin{tabular}{lccc}
\toprule
\textbf{{Workload}} & \textbf{{TTFT MAPE}} & \textbf{{TPOT MAPE}} & \textbf{{E2E MAPE}} \\
\midrule
{CD} & {1.7\%} & {2.1\%} & {1.5\%} \\
{CA} & {2.2\%} & {2.5\%} & {1.9\%} \\
{SUM} & {2.4\%} & {2.8\%} & {2.2\%} \\
\bottomrule
\end{tabular}}

\end{table}

\subsubsection{{Comparison with Model-based Control Alternatives}}
{Table~\ref{tab:control_alternatives} compares ETCInfer with \textit{Greedy threshold}, \textit{Deterministic optimization}, \textit{MPC}, and \textit{Robust MPC+filter}. All share telemetry, actions, SLOs, and R2/V2 traces; results are averaged across scenarios and traces.}
{Model-based baselines improve over greedy thresholding, but ETCInfer gives the best energy--SLO--thermal balance. Deterministic optimization is brittle under state-estimation error. MPC improves stability but degrades when hidden thermal states or token latency drift from the calibrated model. Robust MPC+filter reduces violations but conservative. ETCInfer performs better as its belief encoder and learned world model adapt from observation/action history, while the safety layer rejects risky actions.}

\subsubsection{{Sensitivity to Inference Model Family}}
{We test model sensitivity on Chat Dialog in Table~\ref{tab:model_family_sensitivity}, using locally quantized \textit{Qwen2.5-Instruct}~\cite{qwen2024qwen25}, \textit{DeepSeek-R1-Distill-Qwen}~\cite{deepseekai2025r1}, and \textit{Mistral-Instruct}~\cite{jiang2023mistral}. Workload traces, SLOs, thermal scenarios, telemetry, and scheduler code are unchanged; only model architecture and token latency differ.}
{ETCInfer preserves the Llama-family trend: it reduces energy and throttling while keeping SLO violations below 0.7\%. Thus, its gains depend on telemetry, token-latency estimates, and thermal response, not one checkpoint.}

\subsubsection{Accuracy of Prediction Telemetry}
We validate thermal and power predictors against measured traces. For R2 and V2, {die-temperature RMSE stays below 2.2$^\circ\text{C}$}, {inlet-temperature RMSE below 1.2$^\circ\text{C}$}, {power RMSE below 4.1\%}, and {time-to-throttle error within 5.5\%} for most jobs.
{We validate the calibrated prefill/decode latency estimator on held-out workload traces. Table~\ref{tab:latency_error} reports mean absolute percentage error (MAPE) for TTFT, TPOT, and end-to-end latency. The largest MAPE is 2.8\%, indicating that the estimator is accurate enough to reject actions that threaten SLO constraints. Consistently, Figs.~\ref{fig:slo_ttft}--\ref{fig:slo_tt} show that total SLO violations stay below 0.7\% under high-ambient operation when the estimator guides scheduling decisions.}

\begin{table}[t]
\centering
\caption{{Sensitivity to inference model family on Chat Dialog.}}
\label{tab:model_family_sensitivity}
\resizebox{\linewidth}{!}{
\begin{tabular}{lccc}
\toprule
\textbf{{Inference model family}} &
\textbf{{Energy saving}} &
\textbf{{SLO violation}} &
\textbf{{Throttle reduction}} \\
\midrule
{Qwen2.5-Instruct} & {30.4\%} & {0.43\%} & {88.6\%} \\
{DeepSeek-R1-Distill-Qwen} & {28.7\%} & {0.51\%} & {85.2\%} \\
{Mistral-Instruct} & {26.9\%} & {0.47\%} & {82.4\%} \\
\bottomrule
\end{tabular}}

\end{table}

\section{Related Works}
LLM inference scheduling spans SLO-centric, energy/carbon-aware, and thermal-aware methods, which together motivate coordinated compute--thermal--cooling control.

\textbf{SLO-centric Scheduling.}
SLOs specify latency and availability targets tied to user experience and pricing. SLO-aware schedulers manage admission, batching, and placement using in-flight batching, speculative decoding, and KV-cache optimization~\cite{Jiang2025Efficient}{. Examples include iteration-level scheduling in Orca, prefill/decode disaggregation in DistServe, and chunked-prefill scheduling in Sarathi-Serve~\cite{yu2022orca,zhong2024distserve,agrawal2024sarathi}}. Production systems follow the same goal: vLLM improves throughput with PagedAttention, and TensorRT-LLM improves utilization through kernel fusion and quantization~\cite{li2023vllm,nvidia_tensorrtllm_repo}. These systems sustain throughput and tail latency, but usually treat GPUs as thermally stable and attribute slowdowns to queueing or compute contention. ETCInfer instead uses temperature trends, thermal headroom, and time-to-throttle to guide batching and DVFS under thermal variability.

\textbf{Energy and Carbon-centric Scheduling.}
Another line of work optimizes energy or emissions under latency/deadline constraints~\cite{Wang2022Energy,wang2025storellm}. Examples include data-center right-sizing~\cite{albers2021algorithms}, geo-distributed scheduling with grid coordination~\cite{hu2017coordinating}, low-carbon shifting of delay-tolerant jobs~\cite{wiesner2021let}, adaptive DVFS and power capping~\cite{sun2025learning}, carbon-aware load balancing~\cite{lin2022carbon}, locality-aware data-operator scheduling~\cite{cheng2021network}, and HPC energy plugins~\cite{aaen2023automatic}. These methods reduce energy or emissions while bounding SLO violations, but often assume cooling reacts independently or quickly to power changes. ETCInfer models coupled compute--cooling dynamics and jointly controls ambient setpoint, DVFS, and batching.

\textbf{Thermal-aware Scheduling.}
Thermal-aware scheduling prevents heat-induced performance loss. Prior work shows sustained DNN workloads can overheat GPUs, motivating heuristic and RL schedulers that switch GPU/NPU execution by thermal state~\cite{tan2024thermal}. Automotive SoC studies address CPU--GPU thermal coupling through balanced assignment, co-scheduling, thermal-server abstractions, and utilization bounds~\cite{lee2019thermal,lee2021thermal}. Thermal constraints also affect data centers: {classic studies incorporate cooling cost and spatial thermal effects into placement and consolidation~\cite{moore2005making,pakbaznia2009minimizing}.} Related work studies 3D-stacked LLM memory scheduling~\cite{he2025tasa}, {coordinated cooling and compute management for AI datacenters~\cite{abera2026coordinated}}, and {thermal modeling~\cite{zhang2026hotgpu} and energy-saving techniques for cloud datacenters~\cite{lin2023thermal}}. TAPAS uses historical thermal signals for VM placement and routing~\cite{stojkovic2025tapas}. TAWS integrates voltage/frequency behavior into batching and placement under higher ambient-temperature standards~\cite{lu2025thermal}. ETCInfer further couples ambient setpoints, GPU DVFS, and micro-batch sizing to reduce energy while preserving TTFT, TPOT, and thermal safety.

\section{Conclusion}
In this paper, we present ETCInfer, a thermal-aware scheduler that jointly manages ambient temperature, GPUs, and micro-batch sizing to minimize inference job energy under thermal and latency constraints. By modeling coupled compute–cooling dynamics and leveraging learning-based control, ETCInfer adaptively selects safe setpoints and schedules in response to workload and thermal dynamics. Extensive experiments across real-trace simulations show that ETCInfer significantly reduces total energy, thermal throttling, and SLO violations compared to SOTA baselines. Our results emphasize the need for coordinated control in future AI datacenters and highlight the potential of physics-informed learning in achieving energy-efficient, thermally safe LLM inference.

\bibliographystyle{ieeetr}
\bibliography{ref.bib}

\end{document}